\documentclass[nofootinbib,amsmath,amssymb,showpacs,showkeys,aps]{revtex4-1}
\usepackage[utf8,latin1]{inputenc}
\usepackage{graphicx}
\usepackage{dcolumn}
\usepackage[dvipsnames]{xcolor}
\usepackage[T1]{fontenc}

\usepackage{mathrsfs}  
\usepackage{cases}
\usepackage{bm}
\usepackage{academicons}
\usepackage{mathtools, nccmath}
\usepackage{fancyhdr}
\usepackage{tikz,xcolor}
\newcommand{\qm}[1]{``#1''}
\usepackage{tensor}
\usepackage[normalem]{ulem}
\usepackage{lipsum}
\usepackage{soul}
\usepackage{cancel}
\usepackage{stackengine,scalerel}
\usepackage{mathabx}
\usepackage{hyperref}
\usepackage{tabularx}
\hypersetup{colorlinks, linkcolor={red},citecolor={blue},urlcolor={blue}}

\definecolor{lime}{HTML}{A6CE39}
\DeclareRobustCommand{\orcidicon}{
	\begin{tikzpicture}
	\draw[lime, fill=lime] (0,0) 
	circle [radius=0.16] 
	node[white] {{\fontfamily{qag}\selectfont \tiny ID}};
	\draw[white, fill=white] (-0.0625,0.095) 
	circle [radius=0.007];
	\end{tikzpicture}
	\hspace{-2mm}
}

\newcommand{\dd}{{\rm d}}
\newcommand{\ii}{{\rm i}}
\newcommand{\ee}{{\rm e}}

\fancypagestyle{plain}{%
  \fancyhf{}
  \fancyfoot[C]{\iffloatpage{}{\thepage}}
  }
\foreach \x in {A, ..., Z}{%
	\expandafter\xdef\csname orcid\x\endcsname{\noexpand\href{https://orcid.org/\csname orcidauthor\x\endcsname}{\noexpand\orcidicon}}
}

\begin{document}

\title[Avoiding singularities in Lorentzian-Euclidean black holes: \\ the role of atemporality]{Avoiding singularities in Lorentzian-Euclidean black holes: \\ the role of atemporality}

\author{Salvatore Capozziello\orcidA{}$^{1,2,3}$}
\email{capozziello@na.infn.it}
\author{Silvia De Bianchi\orcidB{}$^{4}$} \email{silvia.debianchi@unimi.it} 
\author{Emmanuele Battista\orcidC{}$^{1,2}$\vspace{0.5cm}}\email{ebattista@na.infn.it} \email{emmanuelebattista@gmail.com}

\affiliation{$^1$ Dipartimento di Fisica ``Ettore Pancini'', Complesso Universitario 
di Monte S. Angelo, Universit\`a degli Studi di Napoli ``Federico II'', Via Cinthia Edificio 6, 80126 Napoli, Italy\\
$^2$ Istituto Nazionale di Fisica Nucleare, Sezione di Napoli, Complesso Universitario 
di Monte S. Angelo, Via Cinthia Edificio 6, 80126 Napoli, Italy
\\
$^3$ Scuola Superiore Meridionale, Largo San Marcellino 10, 80138 Napoli, Italy,\\
$^4$ Dipartimento di  Filosofia,  
Universit\`a degli Studi di Milano,  via Festa del Perdono 7, 20122 Milano, Italy. }

\date{\today} 

\begin{abstract}

We investigate a Schwarzschild metric exhibiting a signature change across the event horizon, which gives rise to what we term a Lorentzian-Euclidean black hole. The resulting geometry is regularized by employing the Hadamard \emph{partie finie} technique, which allows us to prove that the metric represents a solution of vacuum Einstein equations. In this framework, we introduce the concept of \emph{atemporality} as  the dynamical mechanism responsible for the transition from a regime with a real-valued time variable to a new one featuring an imaginary time. We show that this mechanism prevents the occurrence of the  singularity and, by means of the regularized Kretschmann invariant, we discuss in which terms atemporality can be considered as the characteristic feature of this black hole.

\end{abstract}

\maketitle

\section{Introduction}

Signature-changing metrics have been investigated both in quantum and classical general relativity (GR). In quantum cosmology, Lorentzian-signature universes stem out from  Riemann spaces \cite{Gibbons1990,Dereli1993} preventing the appearance of singularities as the result of  spacetime topology fluctuations. 
In this framework, the idea that the metric experiences a signature change at the beginning of cosmological evolution features various approaches to the quantum cosmology problem: $i)$ the Hartle-Hawking no-boundary proposal \cite{Hartle1983,Hawking1993}, where our observed universe is supposed to originate in a regime with no time and no boundary; $ii)$ the Linde picture \cite{Linde1983}, which defines the wave function of the universe via an anti-Wick rotation;  $iii)$ the Vilenkin tunneling from nothing  \cite{Vilenkin1982,Vilenkin1998,Alexandre2023b}, according to which early universes emerge from a primordial quantum tunneling. In general, the signature change is necessary to interpret the solutions of the Wheeler-DeWitt equation related to the geometrodynamics of the so-called superspace, the configuration space of all  three-metrics \cite{Halliwell:1989myn, Capozziello:1999xr}. Metrics undergoing a signature transition have also been studied in loop quantum cosmology  \cite{Mielczarek2012,Bojowald2016,Bojowald2020}, as well as in higher-dimensional models \cite{Embacher1994,Kerner1993,Mohseni2000}, supergravity and string theory \cite{Barrett1993,Mars2000,Stern2018}.

In classical GR, it is usually assumed that the metric keeps its signature unaltered, although this is not ensured by the Einstein equations. Thus, this requirement can be safely loosened \cite{Dray1996a,Dray1996b} and solutions with   signature or even topology changes can be obtained \cite{Horowitz1990}. The Einstein field equations remain, in fact, well-defined since both the Ricci tensor $R_{\mu \nu}$ and the stress-energy tensor $T_{\mu \nu}$ automatically follow the sign change of $g_{\mu \nu}$, whereas the sign of the Ricci scalar $R=g^{\mu \nu} R_{\mu \nu}$ remains unaffected. In this context, homogeneous and isotropic Friedman-Robertson-Walker geometries, representing the classical counterpart of the  models framed in quantum cosmology, have been first investigated in Refs.  \cite{Ellis1992a,Ellis1992b}. For various choices of matter fields, it is shown that the ensuing universes share similar properties with quantum scenarios satisfying the Hartle-Hawking no-boundary condition, as they have an origin in time despite not having a true beginning. Recently, an extended pattern, where all metric components flip their sign simultaneously, has  been  explored in Ref. \cite{Alexandre2023}. 

In the literature, two approaches are usually pursued to describe the behaviour of the temporal component $g_{00}$ of the metric, depending on whether its sign modification is given in terms of a continuous or a discontinuous function, with the latter being considered the more physically meaningful option \cite{Ellis1992a,Ellis1992b}. These choices are set apart by the  singularity nature that unfolds on the hypersurface $\Sigma$ where the metric signature changes. A basic feature is the use of the (Darmois) junction conditions at $\Sigma$ \cite{Israel1966,Barrabes1989,Barrabes-Israel1991,Barrabes1998,Barrabes-book2003,Poisson2002,Poisson2009}, whose fulfillment guarantees that the field equations are satisfied everywhere off $\Sigma$ and go smoothly through it (for a different method, see Refs. \cite{Hayward1993,Hayward1995}). For a timelike or spacelike hypersurface $\Sigma$, junction conditions involve its induced (three-)metric and extrinsic curvature, which are required to be continuous across it. If the latter condition is not satisfied, then the discontinuities in the first-order transverse derivatives of the metric yield a Dirac-delta-like singularity in the Riemann tensor, which can be physically interpreted as the presence of a surface layer (or thin shell) at $\Sigma$ having a distributional stress-energy tensor. When $\Sigma$ is null, both first and second derivatives of the metric can be discontinuous. As a consequence, apart from a surface layer, the hypersurface $\Sigma$ can give rise to a gravitational shock-wave \cite{Poisson2002,Barrabes-book2003}. 

In this paper, we extend the examination of signature-changing metrics in classical GR by considering a Schwarzschild black hole geometry. In our solution,   hereafter referred to as \qm{Lorentzian-Euclidean Schwarzschild black hole}, the metric signature switches from the usual Lorentzian one upon crossing the event horizon. In particular, inside the black hole, the metric admits a Euclidean structure with an ultrahyperbolic signature. The ill-defined distributional terms arising in the Riemann tensor will be treated by borrowing a technique from gravitational-wave theory and compact binaries dynamics, i.e., the Hadamard \emph{partie finie} regularization scheme \cite{Blanchet2000,Blanchet-review(2013),Poisson-Will2014}. In doing so, we will show that neither a surface layer nor a shock-wave is present in our model, since the regularized Riemann tensor (as well as its traceless component, i.e., the Weyl tensor) contains no distributional piece, while the regularized Ricci tensor vanishes everywhere in the spacetime. Moreover, the analysis of the regularized geometry reveals that the velocity of infalling radial trajectories vanishes on the event horizon and becomes imaginary once it is traversed. This means that the singularity at $r=0$ is averted, unlike the standard Lorentzian-signature pattern, where an observer unavoidably reaches it within a finite proper time. 

The Lorentzian-Euclidean Schwarzschild metric can be obtained by considering an imaginary time inside the black hole. In particular, the coordinate time $t$ is naturally replaced by  $it$ as soon as the event horizon is traversed. This is a standard technique widely employed in theoretical physics which is usually referred to as the Wick rotation. The novel facets of our approach are essentially twofold. First of all, we use the Hadamard  method to regularize the metric; then,  we identify the so-called {\it atemporality} \cite{de2023re} as the means explaining why the time variable $t$ becomes imaginary, giving thus a physical meaning to the Wick rotation, which does not represent a mere mathematical trick. Therefore, a crucial aspect of this paper is that we propose atemporality as the dynamical mechanism that allows for the avoidance of the black-hole singularity. Our interpretation of atemporality differs from those presented in current debates on time and timelessness in the foundations of physics, since we do not consider temporality and atemporality as conflicting concepts \cite{braddon2019quantum,anderson2012problem} or, even worse, we do not reject the theoretical and mathematical definition of atemporality as unwanted \cite{Schlegel1948,lam2023laws,mozota2023geometrogenesis}. In addition, our model does not define atemporality simply in terms of the absence of time and its properties, i.e. as the lack of a one-dimensional ordering structure, but as a dynamical process.

The plan of the paper is as follows. After  presenting the Lorentzian-Euclidean Schwarzschild metric in Sec. \ref{Sec:Lorentzian-Euclidean-Schwarzschild-metric}, we consider its regularization process in Sec. \ref{Sec:surface-layer}. Then, we discuss how the atemporality permits to avoid the $r=0$ singularity in Sec. \ref{Sec:avoidance-of-singularity}. Final remarks  are  reported in Sec. \ref{Sec:Conclusion}. Additional material is provided in the appendices.

\section{The Lorentzian-Euclidean Schwarzschild metric} \label{Sec:Lorentzian-Euclidean-Schwarzschild-metric}

As pointed out before, metrics with a varying signature have been analyzed both in quantum settings (in particular, in the context of quantum cosmology \cite{Hartle1983,Dereli1993,Hawking1993}) and at classical level, see e.g. Refs.  \cite{Ellis1992a,Ellis1992b,Alexandre2023}. While many of the existing models are based on the cosmological Friedman-Robertson-Walker geometry, in this section, we present the Schwarzschild metric with a signature-changing behaviour. Such metric assumes the usual Lorentzian-signature form outside the event horizon, it  becomes degenerate at $r=2M$, and then it displays a Euclidean structure for $r<2M$. In this framework, the event horizon constitutes the so-called change surface, i.e., the hypersurface where  the transition between the Lorentzian and Euclidean regimes occurs. 

We first write the  metric in  Schwarzschild coordinates in Sec. \ref{Sec:metric-Schwarzschild-coordinates};  afterwards, we introduce  Gullstrand-Painlev\'e coordinates  in Sec. \ref{Sec:PG-coordinates}. As we will see, this latter analysis will be crucial in Sec. \ref{Sec:surface-layer}, where we will propose a method to regularize the distributional contributions of the Ricci and Weyl tensors. 

Henceforth, we use units $G=c=1$.

\subsection{The Schwarzschild coordinates} \label{Sec:metric-Schwarzschild-coordinates}

In  Schwarzschild coordinates $\{t,r,\theta,\phi\}$,  the Lorentzian-Euclidean Schwarzschild metric takes the form
\begin{align}
    \dd s^2 = - \varepsilon \left( 1-\frac{2M}{r} \right) {\rm d} t^2  
+ \dfrac{{\rm d} r^2 }{\left(1-\frac{2M}{r}\right)} 
+ r^2 \dd \Omega^2,
\label{Lorentzian-Euclidean-Schwarzschild}
\end{align}
where  $\dd \Omega^2 =  {\rm d} \theta^2  + \sin^2 \theta \; 
{\rm d} \phi^2 $ and
\begin{align}
\varepsilon = {\rm sign} \left( 1 - \frac{2M}{r}\right)= 2 H \left( 1 - \frac{2M}{r}\right)-1, 
\label{epsilon-of-r}
\end{align}
the step function $H\left(1-2M/r\right)$ being normalized in such a way that $H(0)=1/2$. As discussed before, the metric undergoes a   signature change as soon as one   crosses the event horizon. Indeed, the function $\varepsilon$  permits to divide the spacetime manifold $V$ into two regions $V_+$ and $V_-$ (i.e, $V = V_+ \cup V_-$) whose common boundary is represented by the change surface  
\begin{align}
\Sigma \; : r=2M,
\label{change-surface}
\end{align}
coinciding with the event horizon. The domain $V_+$  is characterized by the value  $\varepsilon=1$ and pertains to the Lorentzian regime, where the metric is hyperbolic. On $\Sigma$, where $\varepsilon=0$, the metric becomes degenerate by noting that
\begin{align}
{\rm det}\, g_{\mu \nu}=  -\varepsilon \left(r^2 \sin \theta \right)^2.
\label{determinant-metric}
\end{align}
On the other hand,  in $V_-$, where $\varepsilon=-1$, the metric attains an ultrahyperbolic signature and exhibits a Euclidean structure, which means that it has the same features as the Euclidean Schwarzschild metric (recall however that the  Euclidean Schwarzschild solution naturally implements the constraint $r \geq 2M$,  see e.g. Refs. \cite{GH1977,Esposito1992,Battista-Esposito2022,Garnier2024}).  
 
The metric  \eqref{Lorentzian-Euclidean-Schwarzschild} is both divergent and degenerate  on $\Sigma$ as long as we use the Schwarzschild coordinates. Therefore, in order to \qm{disentangle} these two issues,  our analysis  should be performed via a set of coordinates where the only pathologies shown by the metric on the change surface are related to its degeneracy.  We can thus employ either the Kruskal-Szekeres coordinates (see Appendix \ref{App:Kruskal-Szekeres-coordinates}) or the Gullstrand-Painlev\'e ones. As the latter turn out to be more convenient for our purposes,  the ensuing form assumed by the metric  \eqref{Lorentzian-Euclidean-Schwarzschild} will be studied in the next section.

\subsection{ The Gullstrand-Painlev\'e coordinates }\label{Sec:PG-coordinates}

The singular divergent behaviour shown by   the metric \eqref{Lorentzian-Euclidean-Schwarzschild} at $r=2M$ can be eliminated if we introduce the  Gullstrand-Painlev\'e coordinates $(\mathscr{T},r,\theta,\phi)$ \cite{Poisson2009,Blau}, where the parameter $\mathscr{T}$ is, in the Lorentzian domain $V_+$, the proper time measured by a freely falling observer starting off at  rest from infinity and moving radially inward. The four-velocity of such observer can be defined by supposing that, in the spacetime manifold $V$, there exists a family of fundamental worldlines having tangent vector
\begin{align}
u^\mu = \frac{\dd x^\mu}{  \dd \sigma}:= \dot{x}^\mu,
\label{u-mu-d-sigma} 
\end{align}
which, following the recipe of Refs. \cite{Ellis1992a,Ellis1992b}, satisfies
\begin{align}
g_{\mu \nu} u^\mu u^\nu = - \varepsilon. 
\label{velocity-norm}
\end{align}
Here, $\sigma$  is an affine parameter along the worldlines which  corresponds to the proper time $\tau = \int \sqrt{-\dd s^2}$ in the Lorentzian regime \footnote{In the models where the metric is positive-definite in $V_-$, the parameter $\sigma$ can coincide with the proper distance $d = \int \sqrt{ \dd s^2}$ in the Euclidean regime \cite{Ellis1992a,Ellis1992b}.} and it is constructed in such a way to be continuous across $\Sigma$ . 

Starting from Eq. \eqref{velocity-norm}, it is easy to show that the geodesic equation, evaluated in the equatorial plane $\theta=\pi/2$,  is given by
\begin{align}
\dot{r}^2 + \left(1-\frac{2M}{r}\right) \left(\frac{L^2}{r^2} + \varepsilon \right)= \frac{E^2}{\varepsilon^3},
\label{geodesic-equation}
\end{align}
where 
\begin{align}
E&= \varepsilon^2 \left(1-\frac{2M}{r}\right) \dot{t},
\label{energy-PG}
\\
L&= r^2 \dot{\phi},
\label{angular-momentum-PG}
\end{align}
are the constants of motion associated with the static and rotational Killing vectors, respectively.    We  note that, coherently with Eq. \eqref{velocity-norm}, the energy $E$ vanishes on the change surface. When $\varepsilon=1$,  Eq. \eqref{geodesic-equation} describes timelike geodesics in the usual Lorentzian-signature Schwarzschild geometry, while, for $\varepsilon=-1$,  the geodesic equation assumes a form which is consistent with the  features of the Euclidean Schwarzschild geometry (see Ref. \cite{Battista-Esposito2022} for a detailed study of the geodesic motion framed in  the Euclidean Schwarzschild spacetime). An observer whose motion starts at rest from infinity and proceeds radially has $E=\varepsilon^2$ and $L=0$. Then, Eq.  \eqref{geodesic-equation} gives
\begin{align}
\dot{r}=-\sqrt{\varepsilon} \sqrt{\frac{2M}{r}},   
\label{dot-r-PG}
\end{align}
which jointly with  Eq. \eqref{energy-PG}  yields
\begin{align}
\frac{\dd t }{\dd r}= \frac{\dot{t}}{\dot{r}}= -\frac{1}{\sqrt{\varepsilon}}  \sqrt{\frac{r}{2M}} \left(\frac{1}{1-2M/r} \right).    
\label{dot-t-PG}
\end{align}
Upon solving  Eqs. \eqref{dot-r-PG} and \eqref{dot-t-PG}, we obtain 
\begin{align}
r(\sigma) &= \left(\frac{3 \sqrt{2M}}{2}\right)^{2/3} \left(\sigma_0 - \sqrt{\varepsilon}\sigma\right)^{2/3},
\label{r-of-sigma-PG}
\\
\sqrt{\varepsilon} \, t(r) & = t_0 -\frac{2 r^{3/2}}{3 \sqrt{2M}} - \Theta(r),
\label{t-of-r-PG}
\end{align}
where $\sigma_0$ and $t_0$ are integration constants, and
\begin{align}
\Theta(r):=    2 \sqrt{2M} \left[\sqrt{r}-\sqrt{2M} \, {\rm arccotgh} \left(\sqrt{2M/r}\right) \right],
\end{align}
is the solution of the differential equation
\begin{align}
\frac{\dd \Theta}{\dd r} = \frac{1}{1-2M/r} \sqrt{\frac{2M}{r}}.     
\label{d-theta-d-r-PG}
\end{align}
Eqs. \eqref{r-of-sigma-PG} and \eqref{t-of-r-PG} suggest to  define the parameter $\mathscr{T}$ via 
\begin{align}
\sqrt{\varepsilon} \, \mathscr{T} (t,r) :=  \sqrt{\varepsilon} \, t  + \Theta (r),
\end{align}
and hence  it follows from Eq. \eqref{d-theta-d-r-PG} that
\begin{align}
\sqrt{\varepsilon} \, \dd \mathscr{T} =  \sqrt{\varepsilon} \,  \dd t + \frac{\sqrt{2M/r}}{1-2M/r} \dd r,
\end{align}
which implies that   metric \eqref{Lorentzian-Euclidean-Schwarzschild}, written in Gullstrand-Painlev\'e coordinates $(\mathscr{T},r,\theta,\phi)$,  becomes
\begin{align}
\dd s^2 = -\varepsilon \, \dd \mathscr{T}^2 + \left(\dd r + \sqrt{\varepsilon} \sqrt{\frac{2M}{r}} \dd \mathscr{T}\right)^2    + r^2 \dd \Omega^2.
\label{metric-PG-coordinates}
\end{align}
It is worth noticing that formulas underlying the coordinate transformation derived in this section are singular when $r=2M$, akin to the usual Lorentzian-signature pattern, where this behaviour reflects the difficulties  of the Schwarzschild coordinates on the event horizon. In our model, this feature is also ascribed to the  degeneracy-related problems exhibited by the metric on the change surface.

Thanks to the  Gullstrand-Painlev\'e coordinates, the metric remains finite at $r=2M$, and we can formalize its degenerate character  as follows. Let us describe the jump discontinuity of any tensorial quantity $F$ across $\Sigma$ via the notation \cite{Israel1966,Barrabes1989,Barrabes-Israel1991,Ellis1992a,Ellis1992b,Barrabes-book2003,Poisson2009}
\begin{align}
 \left[F\right]:= F \vert_+ - F \vert_-,    
\label{jump-discontinuity-definition}
\end{align}
where $F \vert_+$ (resp. $F \vert_-$) is the limit of $F$ as the point in $V_+$ (resp. $V_-$) approaches $\Sigma$. It then follows from Eq. \eqref{metric-PG-coordinates} that both the metric tensor and its first order derivatives are  discontinuous across $\Sigma$, i.e., 
\begin{align}
\left[ g_{\alpha \beta}\right] &\neq 0,    
\label{metric-discontinuities}
\\
\left[\partial_\mu g_{\alpha \beta}\right] &\neq 0.
\label{derivative metric-discontinuities}
\end{align}
On the other hand,  the induced two-metric $\sigma_{A B}$ $(A=2,3)$ on $\Sigma$ is continuous as 
\begin{align}
\left[\sigma_{A B}\right] &= 0,  
\end{align}
which indicates that the change surface has a well-defined intrinsic geometry \footnote{Recall that the induced line element on $\Sigma$ reads as $\dd s^2_\Sigma = \left(2M\right)^2 \dd \Omega^2$.}. 

Discontinuities \eqref{metric-discontinuities} and \eqref{derivative metric-discontinuities} entail the presence of a distribution-valued Riemann tensor $R^\alpha_{\; \beta \mu \nu}$. In particular, Dirac-delta terms can be present  both in the Ricci and  Weyl parts of $R^\alpha_{\; \beta \mu \nu}$ \cite{Barrabes-Israel1991,Barrabes-book2003,Poisson2009}. The former can be physically associated with the occurrence of a thin distribution of matter at $\Sigma$, referred to as surface layer or thin shell; the latter, describe an impulsive gravitational wave, i.e., a short but violent burst of gravitational radiation. These aspects will be dealt with in the next section.

\section{The regularization process} \label{Sec:surface-layer}

In many GR applications, one has to deal with the problem of joining two  metrics at a common boundary, represented by an embedded hypersurface which divides the spacetime into two distinct regions. This question has been first investigated within the well-known Israel formalism, where the  metric is supposed to keep its signature unchanged, and the hypersurface, say $\bar{\Sigma}$, is either timelike or spacelike \cite{Israel1966,Barrabes1989,Poisson2009}. In this setup, after having expressed the combined metric $g_{\alpha \beta}$ in a set of coordinates covering both sides of $\bar{\Sigma}$, it is demonstrated that even if $g_{\alpha \beta}$  is assumed to be continuous across the hypersurface  (i.e.,  it satisfies  $\left[g_{\alpha \beta}\right]=0$), its first derivatives might exhibit discontinuities  in the direction normal to it. These cause the Riemann tensor to contain Dirac-delta contributions having support on $\bar{\Sigma}$, and,   as a consequence,  $\delta$ quantities will emerge in the Einstein field equations. These distributional factors, in turn, give rise to  a  surface layer whose  stress-energy tensor can be characterized in terms of the discontinuities  across  $\bar{\Sigma}$ of the  extrinsic curvature.

When $\bar{\Sigma}$ is either timelike or spacelike, only the Ricci part of the curvature tensor can possess a distributional singularity. On the other hand, in the case of null hypersurfaces  \cite{Barrabes1989,Barrabes-Israel1991,Poisson2002,Barrabes-book2003} both the Ricci and Weyl tensors can present Dirac-delta singularities. Like before, the former generate a surface layer with a stress-energy tensor depending on the geometric properties of the hypersurface (in particular, it can  be expressed through the jump discontinuities across $\bar{\Sigma}$ of the transverse curvature three-tensor, see Refs. \cite{Poisson2002,Poisson2009} for further details). Moreover, the $\delta$ component of  the Weyl tensor consists, in general, of a Petrov type N part and a Petrov type II piece, and it can be interpreted as an impulsive gravitational wave \cite{Barrabes-book2003} (see
e.g. Refs.\cite{Hotta1992,Esposito2006,Battista_Riemann_boosted,Ortaggio2018,Podolsky2019,Alexander2023,Samann2023} for some investigations on the topic of impulsive waves exploiting either the boosting procedure or the Penrose \qm{cut-and-paste} method).

When signature-varying metrics are considered, a crucial novel facet comes out. Indeed, upon writing the metric in a system of coordinates covering both sides of the change surface $\Sigma$, one finds that $g_{\alpha \beta}$ is discontinuous owing  to its degenerate nature, i.e.,  it inexorably fulfills the condition $\left[g_{\alpha \beta}\right] \neq 0$. In general, this  feature can point out the existence of a thin shell at $\Sigma$. However, in the literature, there exist models where the occurrence of the surface layer can be prevented.  As discussed before, in the context of  Friedman-Robertson-Walker geometries, it has been shown  that metrics having a changing signature  can represent classical solutions of Einstein equations which evolve smoothly through $\Sigma$  \cite{Ellis1992a,Ellis1992b}. In this analysis,  the junction conditions which permit to avoid the formation of the thin shell are satisfied by exploiting the properties of some material stress-energy tensor defined in the spacetime. However, the same approach cannot be pursued in  our model for two reasons. First of all, the material stress-energy tensor of the Schwarzschild solution is vanishing. Secondly, differently from Refs. \cite{Ellis1992a,Ellis1992b}, the change surface \eqref{change-surface} is a null hypersurface. For these reasons, in the following,  we will propose  a method to regularize the $\delta$-function part of the Riemann tensor, thereby preventing the occurrence of both a surface layer and an impulsive wave at $\Sigma$, see  Sec. \ref{Sec:Regularization-Riemann}. As we will see, our regularization scheme works independently of the eventual matter content of the spacetime, which therefore can be also empty. Then, we conclude the section with a digression on degenerate metrics, see Sec. \ref{Sec:degenerate-metrics}.

\subsection{Avoidance of the surface layer and the impulsive wave}\label{Sec:Regularization-Riemann}

As pointed out before, the Lorentzian-Euclidean Schwarzschild metric, written in the Gullstrand-Painlev\'e coordinates,  does not blow up on the change surface  \eqref{change-surface}, where the only  pathological behaviour is due to its  degeneracy. Therefore, Eq.  \eqref{metric-PG-coordinates} can represent the starting point for our investigation  of the properties of the change surface $\Sigma$. 

Discontinuities  \eqref{metric-discontinuities} and \eqref{derivative metric-discontinuities} indicate, in general,   that $\Sigma$ is a singular null hypersurface, as both the Ricci and Weyl parts of the Riemann tensor can exhibit Dirac-delta-like contributions. These are due to the fact that some components of the curvature tensor depend on the derivatives $\varepsilon^{\prime}$, $\varepsilon^{\prime 2}$, and $\varepsilon^{\prime \prime}$, which, in view of  Eq. \eqref{epsilon-of-r},  result in the presence of both linear and quadratic factors in the Dirac-delta function $\delta\left(r-2M\right)$  (see Appendix \ref{App:Riemann-tensor}; hereafter a prime denotes the derivative of a function with respect to its argument). As a consequence, we obtain an ill-defined expression for $R^\alpha_{\; \beta \mu \nu}$ which we propose to regularize by adopting the following procedure.

\begin{figure}[bht!]
\centering\includegraphics[scale=0.70]{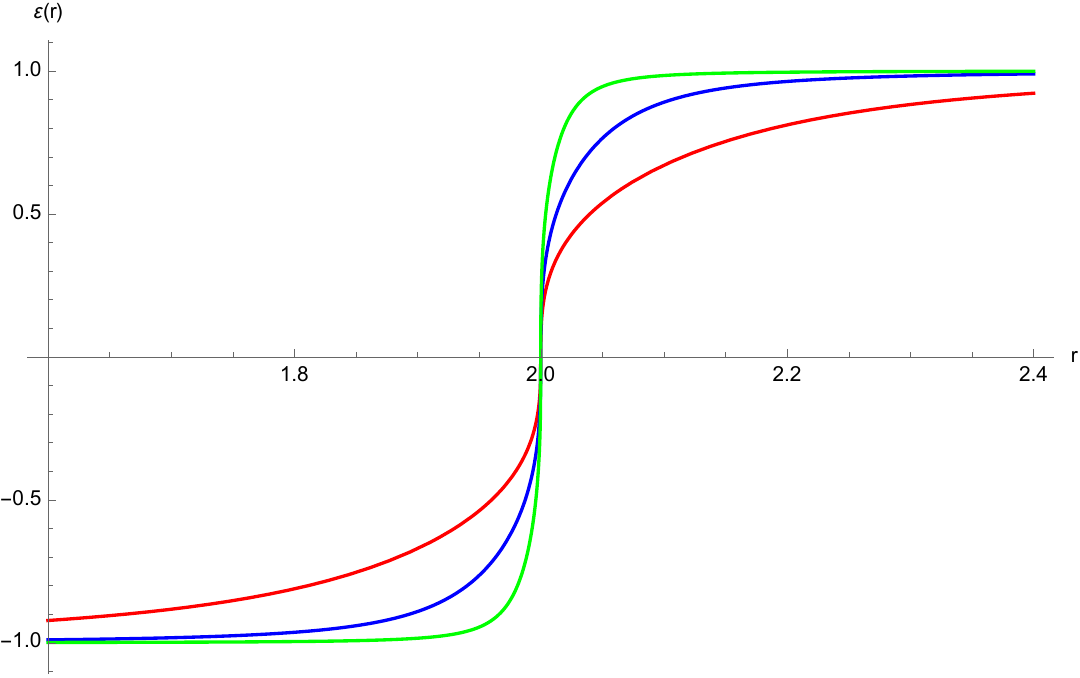}\hspace{1.25cm}
\caption{The approximating function \eqref{sign-regularization} with  $M=1$, $\kappa=1$, and  $\rho=0.1 \, M^2$ (red curve), $\rho=0.01 \, M^2$ (blue curve), and $\rho=0.001 \, M^2$ (green curve). It is clear that $\varepsilon(r)$ gets sharper in the neighbourhood of $r=2M$ as $\rho$ approaches zero. }
\label{Fig-approx-epsilon-2}
\end{figure}

\begin{figure}[bht!]
\centering\includegraphics[scale=0.70]{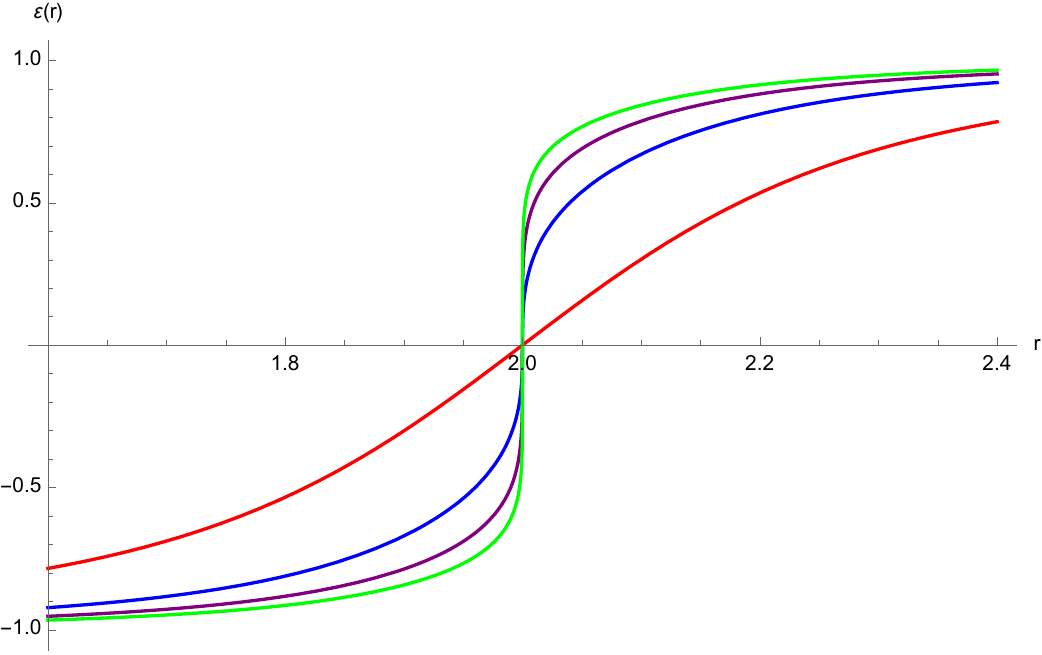}\hspace{1.25cm}
\caption{The approximating function \eqref{sign-regularization} with  $M=1$, $\rho=0.1 \, M^2$, and $\kappa=0$ (red curve), $\kappa=1$ (blue curve), $\kappa=2$ (purple curve), and $\kappa=3$ (green curve). The plot shows that larger values of $\kappa$ result in a steeper function around $r=2M$. }
\label{Fig-approx-epsilon}
\end{figure}

First of all, we introduce a smooth approximation of the sign function, where Eq. \eqref{epsilon-of-r} is replaced by
\begin{align}
\varepsilon(r) = \frac{\left(r-2M\right)^{1/(2 \kappa +1)}}{\left[\left(r-2M\right)^2 + \rho\right]^{1/2(2 \kappa +1)}},
\label{sign-regularization}
\end{align}
$\rho$ being a small positive quantity (having the dimensions of a squared length) and $\kappa$ a  positive integer. These parameters regulate the behaviour of the function around $r=2M$: the smaller is the value assumed by $\rho$, the sharper is $\varepsilon(r)$; in addition, the larger is $\kappa$, the steeper is  $\varepsilon (r)$ (see Figs. \ref{Fig-approx-epsilon-2} and \ref{Fig-approx-epsilon}). Our analysis demonstrates that a well-defined expression of the curvature tensor requires $\kappa \geq 1$ (as we will see below,  this hypothesis is necessary to regularize the $\delta^2$ terms arising in some components of $R^{\alpha}_{\;\beta \mu \nu }$). 

To preserve the  distributional nature of the curvature tensor on $\Sigma$, in our regularization process the approximation \eqref{sign-regularization} is not applied to the derivatives of $\varepsilon(r)$. In this way, $\varepsilon^{\prime}$ and $\varepsilon^{\prime \prime}$  give rise to the Dirac delta and its first-order derivative, respectively. 

Details of the regularization process are reported in Sec. \ref{Sec:regularization-tensors}; then, we further explore the  application  domain of this methodology in Sec. \ref{Sec:Thermodynamics}, where we briefly delve into the thermodynamic properties of  the Lorentzian-Euclidean black hole.

\subsubsection{Regularization of the Riemann, Ricci, and Weyl tensors}\label{Sec:regularization-tensors}

We begin our investigation of the Riemann tensor by considering first the pieces linear in the Dirac-delta function. By recalling that  all the quantities involving the Dirac $\delta$ should be viewed as distributions to be integrated over smooth functions,  we see that our model leads to contributions of the type 
\begin{align}
\int \dd r \; \frac{\delta\left(r-2M\right)}{\varepsilon (r)},
\label{ill-defined-contribution}
\end{align}
which  are in general ill-defined, as  $\delta\left(r-2M\right)/\varepsilon (r)$ is not defined as a distribution, and a naive evaluation  would yield $1/0$ \footnote{We recall that also the product $\varepsilon (r) \delta\left(r-2M\right)$ is not defined as a distribution, since the product between the Dirac delta and the step function is ill-defined. However, the approximation \eqref{sign-regularization} is meant to give a sense also to these terms.}. However, we propose to handle these terms via the approximation \eqref{sign-regularization} jointly with the  Hadamard \emph{partie finie} regularization method \cite{Blanchet2000}, which is a technique broadly employed in the field of gravitational-wave theory and compact binaries dynamics (see Ref.  \cite{Blanchet-review(2013)} and references therein). This   procedure    is based on the following prescription (see Sec. 9.6 in Ref. \cite{Poisson-Will2014} for further details):  
\begin{align}
\frac{\delta(x)}{ \vert x \vert ^n} \equiv 0,
\label{Hadamard-regularization}    
\end{align}
($n$ being a positive integer) and permits to give a meaning to all the ill-defined linear-in-$\delta$ expressions occurring in $R^\alpha_{\; \beta \mu \nu}$, which are seen to vanish in the distributional sense within our model. 

Let us now turn our attention to terms quadratic in the Dirac-delta function, which come up in the components $R^{r}_{\;r \mathscr{T}r}$,  $R^{r}_{\; \mathscr{T} \mathscr{T} r}$, $R^{\mathscr{T}}_{\; r \mathscr{T}  r}$, and  $R^{\mathscr{T}}_{\;  \mathscr{T} \mathscr{T}  r}$  (cf. Eqs.  \eqref{Riemann-1}, \eqref{Riemann-2}, \eqref{Riemann-3}, and \eqref{Riemann-4}). Despite being  ill-defined in the Schwartz theory of distributions \cite{Lieb2001},  these can be readily regularized within our model. This is due to the fact that their coefficients  vanish when $r=2M$ and hence  they amount to zero when integrated over smooth functions.  Therefore,  all the factors involving $\delta^2\left(r-2M\right)$ give no contribution to the Riemann tensor in the distributional sense. It is worth noting that analogous   issues arise also in other research areas, such as  in the context of gravitational shock-waves, where they are addressed in a similar manner (see e.g. Refs. \cite{Dray1984,Sfetsos1995,Battista_Riemann_boosted}).

To show the effectiveness of the above arguments, we will now illustrate how the regularization process can be applied to the component $R^{r}_{\;r \mathscr{T}r}$, which comprises both  $\delta$ and $\delta^2$ terms.  Let $x:=r-2M$, then it follows from Eq. \eqref{Riemann-1} together with the approximation scheme \eqref{sign-regularization}, that the piece linear in $\varepsilon^{\prime}(x)$ leads to an integral proportional to
\begin{align}
    \int \dd x \; \delta(x) \frac{\left(x^2 + \rho\right)^{1/4(2\kappa+1)}}{ x^{1/2(2\kappa+1)}}=  \int \dd x \; \left(\frac{\delta(x) }{x^p x^{1/2(2\kappa+1)}} \right) \left[ x^p \left(x^2 + \rho\right)^{1/4(2\kappa+1)} \right], 
\end{align}
where, in the second step, we have multiplied the integrand by $\frac{ x^p }{ x^p }$ (which is well-behaved in $x=0$),  $p$ being a number such that $p+ 1/2(2 \kappa+1)$ yields a positive even integer. The above integral vanishes, since we can apply the Hadamard regularization scheme \eqref{Hadamard-regularization} to $\delta(x) x^{-p} x^{-1/2(2\kappa+1)} $ and $x^p \left(x^2 + \rho\right)^{1/4(2\kappa+1)}=0$ when $x=0$. The remaining ill-defined quantities appearing in $R^{r}_{\;r \mathscr{T}r}$ stem both from $\varepsilon^{\prime 2}$ and $\varepsilon^{\prime \prime}$ terms. The former yield  a contribution proportional to
\begin{align}
    \int \dd x \;  \frac{x \delta^2(x)}{\varepsilon^{3/2}}  = \int \dd x \;  \delta^2(x) \left(x^2 + \rho\right)^{3/4(2 \kappa+1)} x^{(4 \kappa-1)/2(2\kappa+1)},
\end{align}
which, in the hypothesis $\kappa \geq 1$,  vanishes when evaluated over a smooth function, since the coefficient of  $\delta^2(x)$ is zero in $x=0$. It is thus clear that both the presence of the factor $x$ multiplying $\delta^2(x)$ and  the constraint  $\kappa \geq 1$ are crucial to achieve the desired result. It is worth noticing that an analogous  procedure is exploited  also in the evaluation of the $\delta^2$-part of $R^{\mathscr{T}}_{\; r \mathscr{T}  r}$ and  $R^{\mathscr{T}}_{\;  \mathscr{T} \mathscr{T}  r}$;  the regularization of $R^{r}_{\;  \mathscr{T} \mathscr{T}  r}$ is carried out following similar steps, but, in this case, the condition $\kappa \geq 0$ suffices. Integrals involving  $\varepsilon^{\prime \prime}$  generally  give rise to indefinite $\delta$ and $\delta^2$ quantities which can be easily regularized.  Indeed, in the case of $R^{r}_{\;r \mathscr{T}r}$, we find  an expression of the type
\begin{align}
\int \dd x  \;  \frac{x \varepsilon^{\prime \prime}(x)}{\varepsilon^{1/2}} &= 2\int \dd x \;  \frac{x \delta^\prime(x)}{\varepsilon^{1/2}}
\nonumber \\
&= - 2\int \dd x \;  \delta(x) \frac{\left(x^2 + \rho\right)^{1/4(2\kappa+1)}}{ x^{1/2(2\kappa+1)}} + 2   \int \dd x \;  \delta^2(x) x \frac{\left(x^2 + \rho\right)^{3/4(2\kappa+1)}}{ x^{3/2(2\kappa+1)}}, 
\end{align}
which, as we have explained before, gives zero once we invoke the Hadamard prescription in the linear-in-delta integral and take into account, in the one involving $\delta^2$,   that  $x^{(4 \kappa -1)/2(2 \kappa +1)}=0$ if $x=0$ and $\kappa \geq 1$.

By applying the regularization scheme  to all the indefinite Riemann tensor components involving $\varepsilon^{\prime}$, $\varepsilon^{\prime 2}$, and $\varepsilon^{\prime \prime}$ which are  listed in Eq. \eqref{Riemann-tensor-general-expressions}, we find that these  yield a vanishing contribution in the sense of distributions. Since also the linear-in-delta pieces which require no regularization evaluate to zero,  we can conclude that the Riemann tensor has no distributional behaviour on $\Sigma$, its  final regularized form being given in Eq. \eqref{Riemann-regularized}. However, it should be noted that  the regularized Riemann tensor  has a jump discontinuity across $\Sigma$ due to its unavoidable dependence on $\varepsilon$, i.e., it satisfies the condition $\left[R^{\alpha}_{\; \beta \mu \nu}\right] \neq 0$. Despite that, the ensuing Ricci tensor, Ricci scalar, and consequently,  Einstein tensor, are well-behaved in the whole spacetime, where they are found to vanish. Therefore, there exists no distribution-valued stress-energy tensor associated to the change surface $\Sigma$, which thus does not constitute a surface layer.  Furthermore, our regularization procedure allows us to obtain another  crucial result. In fact, we find that   $\Sigma$ does not represent a new curvature singularity of the spacetime, since the regularized curvature tensor \eqref{Riemann-regularized} yields  a Kretschmann invariant having the usual form, i.e.,
\begin{equation}
R_{\alpha \beta \mu \nu} R^{\alpha \beta \mu \nu}=:\mathcal{K} = \frac{48 M^2}{r^6}, 
\label{Kretschmann invariant}
\end{equation}
which shows that the only singularity of the metric is the one arising in $r=0$ (see also Ref. \cite{Cherubini:2002gen} for a discussion on curvature invariants applied to black holes). 

From our analysis, it is also clear that  the Weyl tensor $C^{\alpha}_{\; \beta \mu \nu}$ stemming from the regularized Riemann tensor \eqref{Riemann-regularized} is discontinuous at $\Sigma$ but has no distributional piece. This implies the absence of any impulsive gravitational wave on $\Sigma$. We can thus conclude that, thanks to our regularization technique, the metric \eqref{metric-PG-coordinates} is a  valid distribution-valued solution of Einstein equations, as the geometrical quantities constructed from it can be properly regularized. 

In the approach pursued so far, we have applied our regularization methodology  to the Riemann tensor \eqref{Riemann-tensor-general-expressions} and then  the resulting regularized expression \eqref{Riemann-regularized} has been exploited to evaluate the Ricci and Weyl tensors. In order to provide a double check of our findings, we have verified that our conclusions are valid also if we follow another program, where we apply the regularization scheme directly to the distribution-valued Ricci and Weyl tensors. These can be derived from Eq. \eqref{Riemann-tensor-general-expressions}, and  the former reads as
\begin{subequations}
\label{Ricci-distribution}
\begin{align}
R_{rr} &= \frac{r (r-2 M) \varepsilon^{\prime^2}+2 \varepsilon  \left[r (2 M-r) \varepsilon^{\prime \prime} +M \varepsilon^\prime \right]}{4 r^2 \varepsilon^2},   
\\
R_{\theta \theta} &= \frac{(2 M-r) \varepsilon^\prime}{2 \varepsilon},
\\
R_{\phi \phi} &= \sin^2 \theta \, R_{\theta \theta},
\\
R_{\mathscr{T} r} &= \sqrt{\frac{2M}{r}} \left(\sqrt{\varepsilon} R_{rr} -\frac{\varepsilon^\prime}{r \sqrt{\varepsilon}}\right),
\\
R_{\mathscr{T} \mathscr{T} } &= \frac{2M-r}{r} \left(\varepsilon R_{rr}-\frac{\varepsilon^\prime}{r}\right),
\end{align}
\end{subequations}
while the nontrivial components of the latter are
\begin{subequations}
\label{Weyl-distribution}
\begin{align}
C^{r}_{\; r \mathscr{T} r} &= - \sqrt{\frac{M}{r}} \frac{ r^2 (r-2 M) \varepsilon^{\prime 2}+2 r \varepsilon  \left[r (2 M-r) \varepsilon^{\prime \prime} +(r-5 M) \varepsilon^\prime\right]+24 M \varepsilon^2}{6 \sqrt{2} r^3 \varepsilon^{3/2}},
\\
C^{r}_{\; \theta \theta r} &= \frac{r (r-2 M) \varepsilon^{\prime 2}}{24 \varepsilon^2}+\frac{r (2 M-r) \varepsilon^{\prime \prime} +(r-5 M) \varepsilon^\prime}{12 \varepsilon }+\frac{M}{r},
\\
C^{r}_{\; \phi \phi r} &= -\left(\frac{r \sin \theta}{2}\right)^2 \sqrt{\frac{2r}{M \varepsilon}} C^{r}_{\; r \mathscr{T} r},
\\
C^{r}_{\; \mathscr{T} \mathscr{T} r} &= \left(\frac{2M-r}{r}\right)  \sqrt{\frac{\varepsilon r}{2M}} C^{r}_{\; r \mathscr{T} r},
\end{align}    
\end{subequations}
the other nonvanishing components being proportional to the ones given above. Also in this case, we find, from Eqs. \eqref{Ricci-distribution} and \eqref{Weyl-distribution},  that  the $\delta$-function part of both $R_{\mu \nu}$ and $C^{\alpha}_{\; \beta \mu \nu}$ is zero in the distributional sense  \footnote{Equation \eqref{Ricci-distribution} yields  the Ricci scalar 
\begin{align*}
R= \frac{r (r-2 M) \varepsilon^{\prime 2}+2 \varepsilon  \left[r (2 M-r) \varepsilon^{\prime \prime}+(M-2 r) \varepsilon^\prime \right]}{2 r^2 \varepsilon^2}.     
\end{align*} 
This can be easily handled via our method, and  the ensuing regularized expression vanishes.}. Like before, the regularized Ricci tensor vanishes, while the regularized Weyl tensor depends on $\varepsilon$ and hence is discontinuous at $\Sigma$.  
Therefore, as pointed out before, we can claim that Eq. \eqref{metric-PG-coordinates} represents a true solution  of the Einstein equations, which thus allow for a  transition between the domains $V_+$ and $V_-$ featuring a non-singular change surface.

\subsubsection{Hawking temperature and black hole entropy}\label{Sec:Thermodynamics}

One of the key features of the regularization procedure set out in the previous section is represented by the approximation \eqref{sign-regularization}. In this section, we show that our approach can be advantageously exploited also to evaluate the otherwise ill-defined Hawking temperature and entropy of the Lorentzian-Euclidean black hole. 

The Hawking temperature is defined through the surface gravity $\kappa_{\rm g}$ as  (we set $\hbar=k_B=1$)
\begin{align}
T_H = \frac{\kappa_{\rm g}}{2\pi},
\end{align}
where, since we are dealing with a spherically symmetric geometry, we have (see e.g. Refs. \cite{Sarkar2007,DelPiano2023} for the standard GR formulas)
\begin{align}
\kappa^2_{\rm g} = \left[-\frac{1}{2 \varepsilon} \left(\nabla_\mu \xi_\nu\right) \left(\nabla^\mu \xi^\nu\right) \right]_{r=2M}=\left[ -\frac{1}{4 \varepsilon} \frac{1}{g_{rr} g_{tt}}   \left(g_{tt}^\prime\right)^2 \right]_{r=2M},
\end{align}
$\xi^\mu = \delta^{\mu 0}$ being the static Killing vector (cf. Eq. \eqref{energy-PG}). It follows, from Eq. \eqref{Lorentzian-Euclidean-Schwarzschild}, that the preceding equation yields
\begin{align}
\kappa_{\rm g} = \left[\frac{\delta\left(1-2M/r\right)}{\varepsilon(r)} \left(1-2M/r\right) + \frac{M}{r^2}  \right]_{r=2M}.
\label{surface-gravity-1}
\end{align}
The first term occurring in the above expression is ill-defined in the theory of distributions (cf. Eq. \eqref{ill-defined-contribution}). However, we can give it a meaning  by exploiting Eq. \eqref{sign-regularization}, which permits to write
\begin{align}
\int \dd r \;  \frac{\delta\left(1-2M/r\right)}{r}  \left(r-2M\right)^{(2 \kappa)/(2 \kappa+1)}  \left[\left(r-2M\right)^2 + \rho\right]^{1/2(2 \kappa +1)}=0,
\end{align}
where we recall that $\kappa \geq 1$. Therefore, Eq. \eqref{surface-gravity-1} gives the usual relation
\begin{align}
\kappa_{\rm g} =  \left(4M\right)^{-1},     
\end{align}
which, in turn, leads to the standard result
\begin{align}
T_H = \left(8 \pi M\right)^{-1}.   
\end{align}
If we now use the first law of thermodynamics $\dd M = T_H \dd \mathcal{S}$, we then obtain the well-known expression for the entropy
\begin{align}
\mathcal{S} = \int \frac{\dd M}{T_H(M)}= 4 \pi M^2= \frac{\mathcal{A}}{4},
\end{align}
where $\mathcal{A}$ is the area of the event horizon. Therefore, our regularization process allows to conclude  that the general formula pertaining to the entropy of (stationary) black holes \cite{GH1977,Iyer1994} is valid also for the Lorentzian-Euclidean one.

Results provided in this section can be the starting point for the investigation of quantum features of the Lorentzian-Euclidean black hole. This examination can be performed along the same lines suggested in Ref. \cite{Hawking1981}. However, the abovementioned similarities between GR and the present model point out that the behaviour of quantum fields around the Lorentzian-Euclidean black hole are essentially the same as in GR, at least outside the black hole. In a forthcoming paper, these topics will be developed in detail.

\subsection{Degenerate metrics} \label{Sec:degenerate-metrics}

In the model devised in this paper, we have supposed that the Lorentzian-Euclidean metric \eqref{metric-PG-coordinates}, or equivalently \eqref{Lorentzian-Euclidean-Schwarzschild}, is valid everywhere in the spacetime. This implies that we allow for a degenerate metric on the change surface \eqref{change-surface}. 

Having a degenerate metric at some points, say $\{\mathscr{P}^\mu\}$, engenders, in general, novel important features which deserve careful consideration.  First of all,   at these points, the metric is not invertible. Therefore, all the equations of interest where we need to eliminate  $g_{\mu \nu}$ by multiplication with its inverse  $g^{\mu \nu}$, are underdetermined, i.e., they may have multiple solutions or no solutions at all. Moreover, some components of $g^{\mu \nu}$ diverge in the neighborhood of $\mathscr{P}^\mu$ and hence other quantities, like the Christoffel symbols, become infinite, too. Furthermore, the equivalence principle is violated as the statement ${\rm det} \, g_{\mu \nu}=0$ is coordinate invariant, which means that no local inertial coordinate system, where the metric takes the Minkowskian form, can be found. Despite these shortcomings, degenerate metrics have been employed in  literature  due to their interesting implications, see    e.g.   Refs. \cite{Horowitz1990,Klinkhamer2013b,Klinkhamer2019a,Klinkhamer2019b,Klinkhamer-Wang2019,Battista2020,Wang2021} and references therein. In these frameworks, one considers extended versions of GR where degenerate metrics are admitted. A first extension can consist in evaluating  relevant quantities at $\mathscr{P}^\mu$  by calculating their limits as $x^\mu$ tends to $\mathscr{P}^\mu$. Another method entails the replacement of $g^{\mu \nu}$ by a generalized notion of inverse metric known as pseudo-inverse, which is derived through the Moore-Penrose algorithm \cite{Penrose1955}. For further details regarding the extensions of GR involving degenerate metrics we refer the reader to Ref. \cite{Gunther}.

Although the Lorentzian-Euclidean metric is degenerate for $r=2M$, we have seen that it is possible to regularize the Riemann tensor and obtain a well-behaved Einstein tensor, which amounts to zero. This means that the Lorentzian-Euclidean metric can be seen as a vacuum solution of an extended version of GR where degenerate metrics are considered.

\section{Avoidance of the singularity at $r=0$} \label{Sec:avoidance-of-singularity}

In the previous section, we have  demonstrated that the change surface \eqref{change-surface} does not represent a surface layer and no impulsive wave is generated in the  spacetime. Therefore, we can now  investigate the behaviour of orbits approaching the event horizon. In particular, we will focus our attention on radially infalling trajectories, which, due to their simple structure, permit to figure out some crucial aspects of the Lorentzian-Euclidean black hole geometry (see Sec. \ref{Sec:radial-geodesics}). We conclude the section by discussing  the possible observational and phenomenological implications   of our findings in Sec. \ref{Sec:implications}.

Hereafter, we employ Schwarzschild coordinates $\{t,r,\theta,\phi\}$.

\subsection{The behaviour of radial orbits} \label{Sec:radial-geodesics}

In this section, we delve into the dynamics of bodies radially nearing the event horizon. We begin our investigation  with freely falling particles in Sec. \ref{Sec:freely-falling-observers}; then, we consider  accelerated observers in Sec. \ref{Sec:accelerated-observers}.  

\subsubsection{Geodesic motion} \label{Sec:freely-falling-observers}

Starting from  Eqs. \eqref{u-mu-d-sigma}--\eqref{angular-momentum-PG}, it is easy to show that the equations pertaining to the radial geodesic motion are
\begin{align}
\dot{r}^2 &= \varepsilon \left(\frac{2M}{r}\right) - \varepsilon \left(1-\frac{E^2}{\varepsilon^4}\right),
\label{radial-geod-1}
\\
\dot{t} &= \frac{E}{\varepsilon^2} \left( \frac{1}{1-2M/r}\right).
\label{radial-geod-2}
\end{align}
Let us consider an observer  starting from rest at some \emph{finite} distance $r_i>2M$. Then, from Eq. \eqref{radial-geod-1}, we obtain
\begin{align}
r_i = \frac{2M}{1-E^2},
\label{initial-radial-distance}
\end{align}
which represents a positive-define length  when
\begin{align}
0<E^2<1.    
\label{energy-constraint}
\end{align}

A first important outcome of our model can be readily understood. In fact, it  follows immediately from Eq. \eqref{radial-geod-1} that $\dot{r}$ attains imaginary values as soon as $r<2M$, or equivalently   $\varepsilon=-1$. We can further investigate this point following the standard analysis of Ref. \cite{Chandrasekhar1985}. Thus, we write the radial variable via the relation
\begin{align}
r(\eta)= r_i \cos^2 \left(\eta/2\right),      
\end{align}
where $\eta \in [0,\eta_H]$, $\eta_H < \pi$ being the value \footnote{In standard GR, one usually assumes $\eta \in [0,\pi]$; however, as explained in this section,  radial geodesics cannot enter the black hole horizon.} of $\eta$ when $r=2M$. In this way, one easily finds that  the equations governing infalling radial geodesics take the form
\begin{align}
\dot{r} &= -\sqrt{\frac{\varepsilon^4 \sin^2 \left(\eta/2\right) + E^2 \left[\cos^2 \left(\eta/2\right) - \varepsilon^4\right] }{\varepsilon^3 \cos^2 \left(\eta/2\right)}},
\label{radial-geod-3}
\\
\dot{t}&= \frac{E}{\varepsilon^2} \frac{\cos^2 \left(\eta/2\right)}{\cos^2\left(\eta/2\right)-\left(1-E^2\right)}.
\label{radial-geod-4}
\end{align}
In view of the constraint \eqref{energy-constraint},  we observe  that the radial velocity  \eqref{radial-geod-3} assumes imaginary values when $\varepsilon=-1$. This same conclusion also applies to the following derivatives:
\begin{align}
\frac{\dd \sigma}{\dd \eta} &= \left(\dot{r}\right)^{-1} \frac{\dd r}{\dd \eta }= r_i \sin \left(\eta/2\right) \cos^2 \left(\eta/2\right) \sqrt{\frac{\varepsilon^3}{\varepsilon^4 \sin^2 \left(\eta/2\right) +E^2 \left[\cos^2 \left(\eta/2\right) - \varepsilon^4\right]}},
\label{d-sigma-d-eta}
\\
\frac{\dd t}{ \dd \eta} &= \dot{t} \, \frac{\dd \sigma}{ \dd \eta} = \frac{E }{\varepsilon^2} \frac{r_i \cos^4 \left(\eta/2\right) \sin \left(\eta/2\right)}{\cos^2 \left(\eta/2\right) - \left(1-E^2\right)} \sqrt{\frac{\varepsilon^3}{\varepsilon^4 \sin^2 \left(\eta/2\right) +E^2 \left[\cos^2 \left(\eta/2\right) - \varepsilon^4\right]}},
\label{d-t-d-eta}
\end{align}
which become imaginary  if $r<2M$. 

At this stage, let us consider the motion when the particle gets to the event horizon. When $\varepsilon=0$, one naively obtains, from Eqs. \eqref{radial-geod-1} and \eqref{radial-geod-2}, or equivalently Eqs. \eqref{radial-geod-3} and \eqref{radial-geod-4},  that both $\dot{r}$ and $\dot{t}$ diverge (note that also in the standard Lorentzian-signature pattern $\dot{t}$ blows up at the horizon). However,  the behaviour of $\dot{r}$ is in contrast with the situation in which the motion starts at rest far away from the black hole. Indeed, in this case $E=\varepsilon^2$ and the radial velocity \eqref{radial-geod-1} boils down to Eq.  \eqref{dot-r-PG}, which vanishes on the change surface and becomes imaginary inside the black hole. This issue can be solved if we recall  that, in our model, the energy is defined as $E=-\varepsilon g_{\mu \nu} \xi^\mu u^\nu$, $\xi^\mu$ being the static Killing vector field  (cf.  Eq. \eqref{energy-PG}). This means that, for a given motion having  $2M<r_i<\infty$, we can write $E^2 =  \alpha^2 \varepsilon^4$, where $\alpha^2$ is some positive-definite bounded function owing to Eq. \eqref{energy-constraint}. In this way, one immediately see either from Eq. \eqref{radial-geod-1} or Eq. \eqref{radial-geod-3} that $\dot{r}$ becomes zero on the change surface, like in the scenario having $r_i \gg  2M$.

In light of the above analysis, we can conclude that  infalling particles moving freely in the radial direction have a velocity $\dot{r}$ which vanishes on the event horizon and becomes imaginary after having crossed it. This can be  interpreted as an indication that the singularity at $r=0$ can be evaded because the observer never reaches it. This scenario can be ascribed to the emergence of an imaginary time as soon as $r<2M$. Indeed, one  way to explain the  metric signature change in Eq. \eqref{Lorentzian-Euclidean-Schwarzschild} consists in supposing that the coordinate time $t$  is no longer a real-valued variable inside the black hole. We propose to relate this feature to the concept of atemporality, which thus configures, in our model, as the dynamical mechanism which permits to avoid the black-hole singularity.

A crucial aspect of our investigation consists in proving that the observer in  radial free fall takes an infinite amount of  proper time to stop at the event horizon (recall from Sec. \ref{Sec:PG-coordinates} that $\sigma$ coincides with the proper time in the Lorentzian domain). This task can be performed by solving Eq. \eqref{d-sigma-d-eta}, which represents a differential equation with distribution-valued coefficients due to the presence of $\varepsilon$ (see Eq. \eqref{epsilon-of-r}). In general, the resolution of this class of equations is challenging and the explicit solution can be exhibited only in some particular cases (see e.g. Ref. \cite{Albeverio2012} for some applications to quantum mechanics). For our purposes,  it is reasonable to face Eq. \eqref{d-sigma-d-eta} in two ways. The first method relies on  simply regarding $\varepsilon$ as a constant coefficient. In this way,  we obtain, from  Eq. \eqref{d-sigma-d-eta},  
\begin{align}
\frac{\sigma(\eta)}{M}&= -\frac{2\sqrt{\varepsilon^3}\sqrt{E^2 \left(1-\varepsilon^4\right)}}{\left(E^2-1\right)\left(E^2-\varepsilon^4\right)}   +\frac{1}{\sqrt{2}\left(E^2-1\right)}\sqrt{\frac{\varepsilon^3}{\varepsilon^4 + E^2 \left(1-2 \varepsilon^4\right)+\left(E^2-\varepsilon^4\right) \cos \eta}}
\nonumber \\
&\times \left\{ \frac{\left[\varepsilon^4 + E^2 \left(3-4\varepsilon^4\right)\right] \cos \left(\eta/2\right)}{\left(E^2-\varepsilon^4\right)} + \cos \left(3 \eta/2 \right) \right\} 
+\frac{\varepsilon^4 \sqrt{\varepsilon^3}}{2 \left(E^2-\varepsilon^4\right)^{3/2}} 
\nonumber \\
&\times \left\{ \log \left[\sqrt{2} \sqrt{E^2 -\varepsilon^4} \cos\left(\eta/2\right) + \sqrt{E^2 + \varepsilon^4 - 2 E^2 \varepsilon^4 + \left(E^2-\varepsilon^4\right) \cos \eta}\right]^4  \right.
\nonumber \\
& \left. -\log \left[ 4 \left(\sqrt{E^2 -\varepsilon^4} + \sqrt{E^2 \left(1-\varepsilon^4 \right)}\right)^4 \right] \right\},
\end{align}
where we have adopted the initial condition $\sigma(\eta=0)=0$ and we recall that $E^2 = \alpha^2 \varepsilon^4$. The behaviour of $\sigma(\eta)$ at $r=2M$ can be thus inferred by letting $\varepsilon$ approach zero. In particular, we find   $\lim\limits_{\varepsilon \to 0^+} \sigma = + \infty$ and $\lim\limits_{\varepsilon \to 0^-} \sigma = \ii  \infty$, with $\cos (n \, \eta_H)$ finite ($n=3/2,1,1/2$). The first limit indicates that the proper time becomes infinite when $r=2M$, while the second is consistent with the fact that $\sigma$ takes imaginary values for $r<2M$. 

\begin{figure}[bht!]
\centering\includegraphics[scale=0.55]{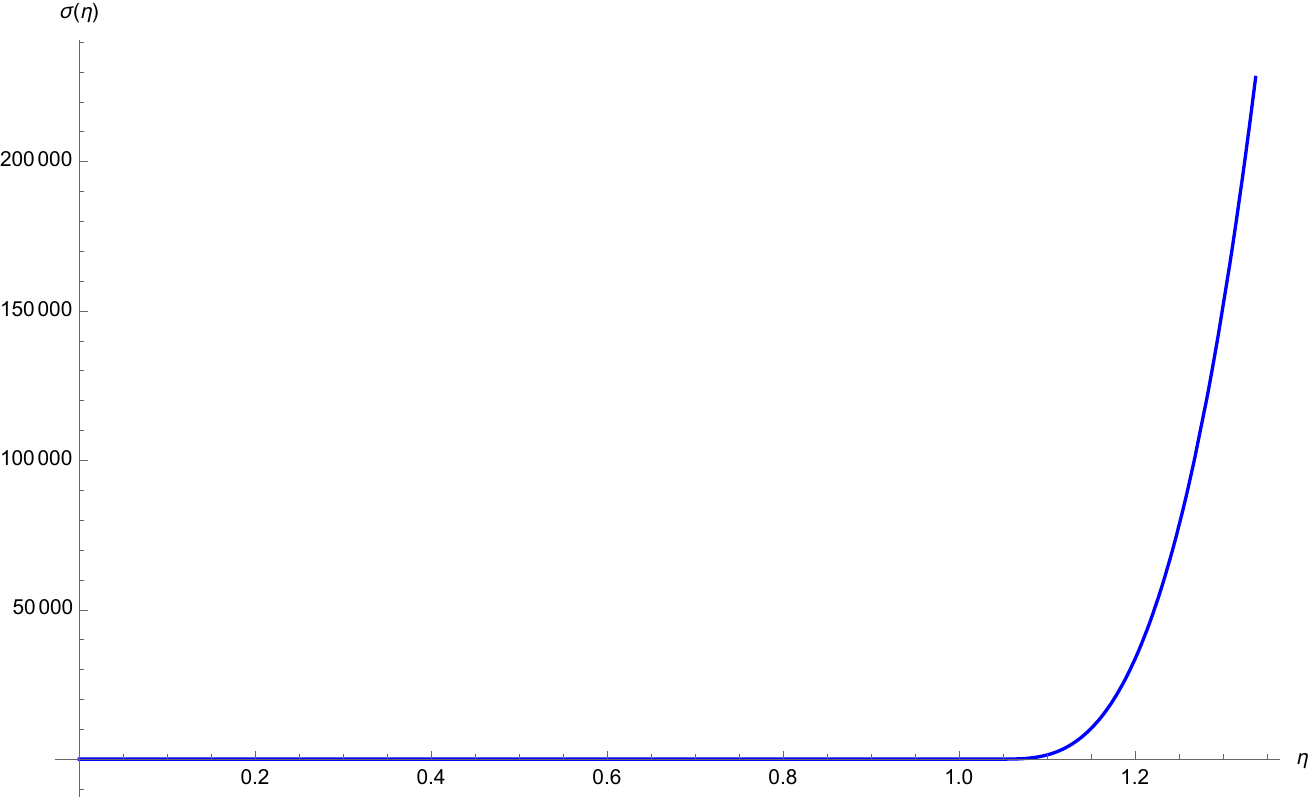}\hspace{1.25cm}
\caption{The solution $\sigma(\eta)$ of the differential equation \eqref{d-sigma-d-eta} obtained via the approximation \eqref{sign-regularization} for $\varepsilon$. The function blows up at the event horizon (which is reached when $\eta \approx 1.3$), meaning that the observer takes an infinite  proper time to reach $r=2M$.}
\label{Fig-sigma-of-eta}
\end{figure}

The second approach consists in solving Eq.  \eqref{d-sigma-d-eta} by adopting the approximation \eqref{sign-regularization} for $\varepsilon$. The ensuing solution, which we have obtained with  numerical methods, attains unboundedly large values when the observer reaches the event horizon, as shown in Fig. \ref{Fig-sigma-of-eta}. 

\begin{figure}[bht!]
\centering\includegraphics[scale=0.55]{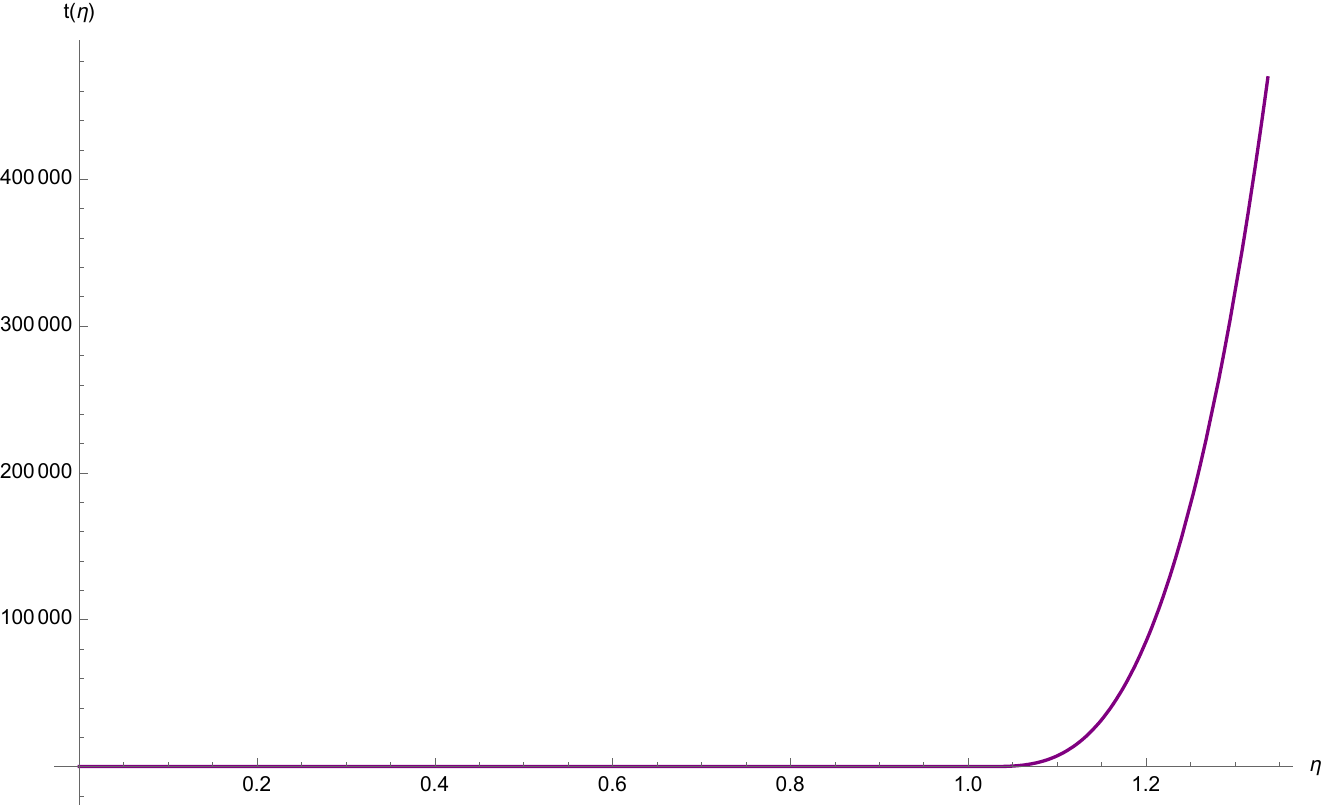}\hspace{1.25cm}
\caption{The solution $t(\eta)$ of the differential equation \eqref{d-t-d-eta} obtained via the approximation \eqref{sign-regularization} for $\varepsilon$. The function diverges at the event horizon (i.e.,  when $\eta \approx 1.3$).}
\label{Fig-t-of-eta}
\end{figure}

By following a similar procedure, we have solved also Eq. \eqref{d-t-d-eta}. As it is clear from Fig. \ref{Fig-t-of-eta}, the coordinate time $t(\eta)$ becomes infinite as the observer gets to the event horizon, in line with the predictions of  the standard Schwarzschild solution. 

At this stage, a final crucial remark is in order. Although  it remains true that the (regularized) Kretschmann invariant $\mathcal{K}$  blows up at $r=0$ (see Eq. \eqref{Kretschmann invariant}), our investigation proves that this point cannot be reached by infalling radial geodesics. In other words, since $2M \leq r(\sigma) < \infty$,  $\mathcal{K}$ is always bounded along the trajectories followed by radial geodesics, as its maximum value is 
\begin{align}
\label{regularK}
\mathcal{K}(r=2M)= \frac{3}{4M^4}.
\end{align}
This ties in with a crucial aspects of singularity theorems \cite{Hawking1970} (see also Ref. \cite{Magalhaes2024} and references therein): there is not necessarily a link between  divergent curvature invariants and  spacetime singularities, the latter being singled out solely by  incomplete (causal) geodesics.

\subsubsection{Accelerated motion} \label{Sec:accelerated-observers}

The dynamics of a  radially accelerated observer, whose trajectory begins at rest from a large distance from the black hole,  is ruled by the equation (see e.g.  Ref. \cite{Formiga2020} and references therein)
\begin{align}
a^\lambda= \frac{\dd U^\lambda}{\dd \sigma} + \Gamma^\lambda_{\mu \nu} U^\mu U^\nu,
\label{acceleration-expr-1}
\end{align}
where $a^\lambda$ denotes the  four-acceleration and 
\begin{align}
U^\mu := \frac{\dd x^\mu}{\dd \sigma},    
\end{align}
the four-velocity. To describe a  radial-directed orbit, we let $\theta$ and $\phi$ be constant in Eq. \eqref{acceleration-expr-1}, and hence we find that the nonvanishing components of $a^\lambda$ are
\begin{align}
a^t &= \frac{\dd U^t}{\dd \sigma} + 2 \Gamma^t_{tr} U^t U^r,
\label{acceleration-t}
\\
a^r &= \frac{\dd U^r}{\dd \sigma} + \Gamma^r_{tt} U^t U^t + \Gamma^r_{rr} U^r U^r,
\end{align}
where (cf. Eq. \eqref{Lorentzian-Euclidean-Schwarzschild})
\begin{subequations}
\begin{align}
\Gamma^t_{tr} &= \frac{M}{r^2 \left(1-2M/r \right)} + \frac{\varepsilon^\prime}{2 \varepsilon},
\label{Gamma-t-t-r}
\\
 \Gamma^r_{tt}&= \frac{M \varepsilon}{r^2} \left(1-2M/r \right) + \frac{\varepsilon^\prime \left(2M-r \right)^2}{2 r^2},
 \label{Gamma-r-t-t}
 \\
 \Gamma^r_{rr} &= -\frac{M}{r^2 \left(1-2M/r \right)}.
 \end{align}
\end{subequations}
By applying the regularization scheme developed in Sec. \ref{Sec:surface-layer} to Eq. \eqref{Gamma-t-t-r}, and resorting to standard distribution-theory tools to evaluate Eq. \eqref{Gamma-r-t-t}, it is easy to show that the terms proportional to $\varepsilon^\prime$ give a vanishing contribution in the distributional sense. 
In this way,  the preceding formulas  yield
\begin{align}
a^t &= \left(1-2M/r \right)^{-1}\frac{\dd }{\dd \sigma} \left[\left(1-2M/r \right) U^t\right],  
\label{acceleration-t-2}
\\
a^r &= \frac{\dd U^r}{\dd \sigma} + \frac{M \varepsilon}{r^2} \left(1-2M/r \right) U^t U^t -\frac{M}{r^2\left(1-2M/r \right)} U^r U^r.
\end{align}
We can now integrate Eq. \eqref{acceleration-t-2} by setting
\begin{align}
    a^t = \left(\frac{\dd }{\dd \sigma} \log \mathcal{F}(\sigma)\right)U^t,
\end{align}
$\mathcal{F}(\sigma)$  being  a generic function of the affine parameter $\sigma$. We thus obtain 
\begin{align}
\frac{\dd }{\dd \sigma} \log \left(\frac{\mathcal{F}}{1-2M/r}\right) =     \frac{\dd }{\dd \sigma} \log U^t,
\end{align}
which, in turn, leads to
\begin{align}
U^t = \frac{\mathcal{F}}{1-2M/r}. 
\label{accel-velocity-t}
\end{align}
Therefore, from the condition $g_{\mu \nu} U^\mu U^\nu = -\varepsilon$, we easily get the radial component of the observer  four-velocity
\begin{align}
U^r= - \sqrt{\varepsilon} \sqrt{\mathcal{F}^2 - \left(1-2M/r \right)}.    
\label{accel-velocity-r}
\end{align}

It follows from Eqs. \eqref{accel-velocity-t} and \eqref{accel-velocity-r} that the coordinate time satisfies
\begin{align}
\frac{\dd t}{ \dd r} = -\left(\frac{\mathcal{F}}{1-2M/r} \right) \frac{1}{ \sqrt{\varepsilon\left[\mathcal{F}^2 -\left(1-2M/r\right)\right]}},
\end{align}
which, jointly with Eq. \eqref{accel-velocity-t}, permits to write in addition the differential equation for $\sigma$
\begin{align}
\frac{\dd \sigma}{\dd r} = -\frac{1}{\sqrt{\varepsilon\left[\mathcal{F}^2 -\left(1-2M/r\right)\right]}}.  \label{d-sigma-d-r-accelerated}
\end{align}

The above analysis indicates that  dynamics of radially accelerated particles shares important  similarities with the freely falling case. Indeed,  Eq. \eqref{accel-velocity-r} unveils a crucial feature of our model:  supposing that $\mathcal{F}$ remains finite at $r=2M$, the  accelerated observer also halts at the event horizon;  moreover, when $r<2M$, the radial velocity $U^r$ becomes imaginary.
\begin{figure}[bht!]
\centering\includegraphics[scale=0.55]{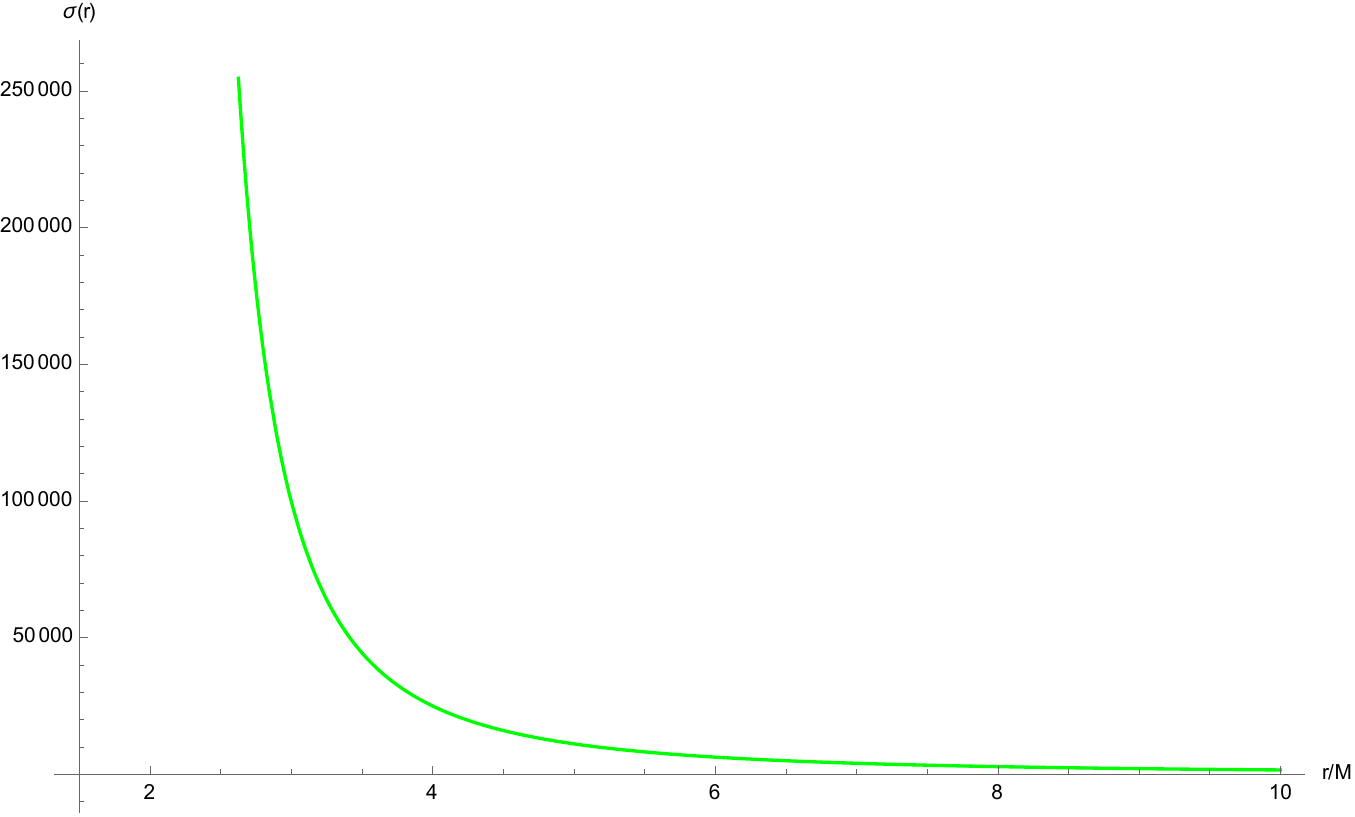}\hspace{1.25cm}
\caption{Proper time $\sigma(r)$ of a radially accelerated observer obtained by numerically solving  the differential equation \eqref{d-sigma-d-r-accelerated}  with the initial condition $\sigma(r_0)=0$. The functions $\varepsilon$ and $\mathcal{F}$ have been evaluated by employing Eqs. \eqref{sign-regularization}   and  \eqref{F-function-a}, respectively. The solution $\sigma(r)$ diverges at the event horizon.  }
\label{Fig-sigma-of-r-accelerated}
\end{figure}

It is clear, from Eqs. \eqref{accel-velocity-t} and \eqref{accel-velocity-r}, that the geodesic orbit of a particle initially at rest at infinity is recovered in the limit $\mathcal{F}=1$ (cf. Eqs. \eqref{energy-PG} and \eqref{dot-r-PG} and recall that such free-fall motion  has $E=\varepsilon^2$). In addition, the choice $\mathcal{F} = \sqrt{1-2M/r}$ pertains to the case of a static observer. Therefore, we can consider, without loss of generality, that the $\mathcal{F}$ function assumes the  form
\begin{align}
 \mathcal{F} = f(\sigma) \sqrt{1-2M/r},   
 \label{F-function-generic}
\end{align}
with $ f(\sigma)  >1$. A particular instance of the model \eqref{F-function-generic} consists in assuming that  $f(\sigma)$ is given by a constant factor $a>1$, so that we can write
\begin{align}
 \mathcal{F} =a \sqrt{1-2M/r}.   
 \label{F-function-a}
\end{align}
By considering the above accelerating term, we have ascertained that both $\sigma(r)$ and $t(r)$  become infinite when $r=2M$ (the function $\sigma(r)$ is displayed in Fig. \ref{Fig-sigma-of-r-accelerated}; the behaviour of $t(r)$ is similar). In other words, like for the geodesic motion, the accelerated body takes an infinite amount of proper time to stop at the event horizon; similarly, a distant observer sees the accelerated particle approaching $r=2M$ in an infinite time.

\subsection{Observational and phenomenological considerations} \label{Sec:implications}

Let us discuss now some  implications  that can characterize our model from a phenomenological standpoint. 

A first scenario that should be taken into account involves the deceleration experienced by bodies  pointing radially towards the event horizon, where  they  ultimately stop. This novel aspect can potentially lead to  some observational signatures of the Lorentzian-Euclidean framework proposed here. However, this possibility comes with some caveats. First of all, as shown in Sec. \ref{Sec:radial-geodesics}, the deceleration  is a long-lasting process which unfolds over an infinite period of time, as seen from both the perspective of  particle and that of an observer stationed at infinity. Thus, the deceleration occurs smoothly rather than in an abrupt way, and this can pose some issues in the search for  possible  detectable properties of the Lorentzian-Euclidean geometry. Moreover, both in the standard GR context and the one studied here, a distant observer never sees the particle crossing the event horizon, meaning that  the resultant measurable effects in the two theories  could be similar. However, the accumulation of a bunch of particles on the event horizon can be a distinguishing feature of our model, which can set it apart from GR. Potentially,  this phenomenon can be an observable feature. The particle accumulation could shape the luminous \emph{silhouette} around the black hole event horizon. The forthcoming observational campaigns of the Event Horizon Telescope collaboration,  which are going to detect the fine structure around supermassive black holes, as recently reported in Refs. \cite{EHT1,EHT2}, could give indications in this direction.

Another point worth mentioning  regards the formation of accretion disks around  black holes \cite{Abramowicz2011}. Let us consider a slowly rotating black hole whose metric can be approximated with the Schwarzschild solution. Since a key aspect of the Lorentzian-Euclidean pattern is that its outer geometry mirrors the Schwarzschild one, the position of stable and unstable circular orbits remains unchanged. Consequently, the black hole is allowed to gather mass, giving rise to a disk with its inner edge located, as usual, near the innermost stable circular orbit. Although  the velocity of radially moving bodies vanishes at the event horizon, matter orbiting the black hole still loses energy as it spirals inward, resulting in the radiation that we observe. Also in this case, the different behaviour of infalling particles with respect to GR might give rise to some observable consequences.

\section{Discussion and conclusions  }
\label{Sec:Conclusion}

In classical GR, it is possible to obtain solutions of Einstein equations whose signature  transitions from the standard Lorentzian to the Euclidean one, and are completely smooth through the surface change. These exhibit interesting properties, as they can be related to the so-called real tunnelling solutions of the Wheeler-DeWitt equation in quantum cosmology \cite{Halliwell1990,Hawking1993}. 

Signature-changing  metrics have been mainly investigated in classical homogeneous and isotropic cosmologies. In these scenarios, a crucial role is fulfilled by the (Darmois) junction conditions, which guarantee the absence of any surface layer at the change surface. In this paper, we have extended this approach by considering  a static and spherically symmetric black hole metric, which gives rise to what we dubbed Lorentzian-Euclidean Schwarzschild black hole. This solution reproduces exactly the standard Schwarzschild geometry for $r>2M$, becomes degenerate at $r=2M$, and exhibits a Euclidean structure when $r<2M$,  see Eqs. \eqref{Lorentzian-Euclidean-Schwarzschild} and \eqref{epsilon-of-r}. In our investigation, we have demonstrated that the event horizon is not a singular hypersurface, as it is not the site  of a surface layer or an impulsive gravitational wave. This result has been achieved thanks to the introduction of a regularization scheme based on the Hadamard \emph{partie finie}  technique, which permits to discard all the distribution-valued ill-defined pieces of both the Riemann and Weyl tensors (see Sec. \ref{Sec:surface-layer}). The resulting regularized tensors depend only on the function $\varepsilon$  and not on its derivatives, while  the regularized Ricci tensor and Ricci scalar vanish. This means that the Lorentzian-Euclidean Schwarzschild metric represents a well-defined signature-changing solution of the vacuum Einstein field equations. 

In our analysis, a crucial role is fulfilled by the concept of atemporality, which describes the Euclidean regime $r<2M$  where the time variable $t$  becomes imaginary. The examination of Sec. \ref{Sec:radial-geodesics} reveals that both infalling radial geodesics and accelerated orbits   have a  velocity which vanishes on the event horizon and becomes imaginary inside the black hole, meaning that the singularity at $r=0$ is not effectively reached. This differs from the standard Lorentzian-signature framework, where  a particle unavoidably falls into the singularity within a finite proper time once it has crossed the event horizon \cite{Wald-book1984}. Therefore, we propose to interpret atemporality as the mechanism which regularizes the black hole geometry and allows to avoid the singularity.  In other words, our model suggests that  objects seem to be unable to enter the event horizon, at least at classical level. The same analysis for quantum particles will be developed elsewhere. Furthermore, as discussed in Sec. \ref{Sec:implications}, the accumulation of a bunch of particles on the event horizon is a feature predicted by our model which differentiates it from GR. This phenomenon can be regarded as a potential observable feature.

Atemporality can be seen as the dynamical procedure introducing a natural cut-off for the radial trajectory $r(\sigma)$, which is subject to the condition $2M \leq r(\sigma) < \infty$, making thus the (regularized) Kretschmann scalar $\mathcal{K}$ bounded along the path of radially-moving observers. Therefore, a crucial point that we want to stress on the ground of Eq. \eqref{regularK} is the following. We can interpret  $\mathcal{K}$ as the quantity related to the implementation of the atemporality mechanism. This parameter permits to establish a relation with the the black hole size (i.e., the larger the black hole, the smaller the value of $\mathcal{K}$ at $r=2M$ and vice versa), and offers  the possibility of \qm{measuring} the degree of atemporality of the system. Indeed, any black hole can be characterized by a mass and, eventually, an angular momentum and a charge. Now we propose to add another parameter: the atemporality, which is related to the regularized Kretschmann invariant \eqref{regularK}. 
In other words, we can state: 
\begin{itemize}
    \item [] \emph{Atemporality is the dynamical mechanism by which an observer pointing towards the event horizon cannot reach the singularity in $r=0$, because  real-valued geodesics and accelerated orbits cannot be prolonged up to there. As a consequence, both the  time variable and radial velocity become imaginary inside the black hole. The parameter \qm{measuring} the degree of atemporality is the Kretschmann scalar $\mathcal{K}$, which is related to the mass of the black hole.}
\end{itemize}

As discussed in this paper, atemporality  means that when time becomes imaginary, there is no dynamics in the standard sense related to the real time. Therefore, we can say that there exists an  analogy with the tunnelling effect in quantum mechanics \cite{Sakurai2011}. Indeed, in  the latter  setup,  the nature of the quantum wave function changes inside and outside the potential barrier. In the present case, the nature of time, as well as that  of  geodesics and accelerated paths, changes in passing through the event horizon, so then  atemporality consists in this change of dynamical behaviour.

Imaginary time is widely employed in quantum gravity and quantum cosmology, where it
is defined via the well-known Wick rotation \cite{Esposito1992} and  offers a unified picture of quantum and statistical mechanics \cite{2005APS..TSS.AB001A}. The outcome of this paper moves towards giving a more physical meaning to Wick rotation, which does not configure as a mere mathematical operation. Indeed, in our framework the emergence of an imaginary time is associated with the transition from the standard Lorentzian geometry to a different one, which is endowed with an ultrahyperbolic signature and is dominated by atemporality. From our perspective, there is no preference between a real-valued or an imaginary time variable,  since we deem that both options can be treated on the same footing and share the same level of importance. This means that the dispute over the fictional status of imaginary time loses  importance \cite{Schlegel1948,deltete1996emerging,de2023re}. In this regard, Hawking himself has stated in his popular science book  (Ref. \cite{Hawking2001-book}, page 59): 
\\
\qm{One might think this means that imaginary numbers are just a mathematical game having nothing to do with the real world. From the viewpoint of positivist philosophy, however, one cannot determine what is real. All one can do is find which mathematical models describe the universe we live in. It turns out that a mathematical model involving imaginary time predicts not only effects we have already observed but also effects we have not been able to measure yet nevertheless believe in for other reasons}.

Atemporality as proposed in this paper can open up new interesting perspectives. First of all, it represents a new original  method to remove the black hole singularity which can be framed in the research field dealing with regularized black hole geometries (see the models studied in e.g. Refs. \cite{Dymnikova1992,Hayward2005,Bambi2013,Platania2019,Eichhorn2021,Ghosh2022,Delaporte2022} and 
the formalism devised in Ref. \cite{Klinkhamer2013}, which has also been generalized to cosmological settings \cite{Klinkhamer-Wang2019,Battista2020,Wang2021}). Therefore, exploring possible connections with these approaches would represent a fascinating topic. Furthermore, the concept of atemporality can enable  new progress in our understanding of black hole physics. In addition, the ideas set out in this paper encourage us to search for possible  observational signatures of atemporality. These  aspects deserve consideration in a separate paper.

\section*{Acknowledgements}
SC and EB acknowledge the support of  Istituto Nazionale di Fisica Nucleare (INFN) Sez. di Napoli, Iniziative Specifiche QGSKY and MOONLIGHT2.  SDB acknowledges the funding and support from the European Union's Horizon 2020 research and innovation program under Grant Agreement No. 758145 -- PROTEUS and the project (2021-0567-COSMOS) funded by Cariplo Foundation.
This paper is based upon work from COST Action CA21136 {\it Addressing observational tensions in cosmology with systematics  and fundamental physics} (CosmoVerse) supported by COST (European Cooperation in Science and Technology). We thank the anonymous referee for the helpful comments received which have improved the quality of the manuscript.

\appendix

\section{The Kruskal-Szekeres coordinates} \label{App:Kruskal-Szekeres-coordinates}

In this Appendix, we  work out  the metric \eqref{Lorentzian-Euclidean-Schwarzschild} in Kruskal-Szekeres coordinates. The following  calculations are valid in $V_+$ and $V_-$, where $\varepsilon = \pm 1$ and $\varepsilon^2 = 1$ (cf. Eq. \eqref{epsilon-of-r}). 

Upon adopting null coordinates $(u,v)$,  standard arguments  \cite{Wald-book1984,Carroll2004,Poisson2009} permit to show that the metric \eqref{Lorentzian-Euclidean-Schwarzschild} takes the form 
\begin{align}
    \dd s^2 = -\varepsilon \left(1-\frac{2M}{r}\right) \dd u \, \dd v + r^2 \dd \Omega^2,
\end{align}
where
\begin{align}
u & = t - r^\star,
\nonumber \\
v& = t + r^\star,
\end{align}
the tortoise coordinate $r^\star$ being 
\begin{align}
r^\star= \frac{1}{\sqrt{\varepsilon}} \int \frac{\dd r}{1-2M/r} = \frac{1}{\sqrt{\varepsilon}} \left[r + 2M \log \left(\frac{r}{2M}-1 \right)\right].  
\end{align}
In this coordinate system, the surface $r=2M$ still represents a coordinate singularity. This can be removed by means of a new set of null coordinates
\begin{align}
u^\prime &= - \varepsilon \, {\rm e}^{-\sqrt{\varepsilon}u /(4M)}, 
\nonumber \\
v^\prime &=  {\rm e}^{\sqrt{\varepsilon} v  /(4M)}, 
\end{align}
which, in view of the preceding formulas, give
\begin{align}
u^\prime &= - \varepsilon \sqrt{\frac{r}{2M}-1}  \,  {\rm e}^{\left(r-\sqrt{\varepsilon} t\right) /(4M)}, 
\nonumber \\
v^\prime &= \sqrt{\frac{r}{2M}-1} \, {\rm e}^{\left(r+\sqrt{\varepsilon} t\right) /(4M)}.
\end{align}
In this way, the Lorentzian-Euclidean Schwarzschild metric becomes
\begin{align}
\dd s^2 = -\varepsilon \frac{32 M^3}{r} {\rm e}^{-r/(2M)} \; \dd u^\prime \dd v^\prime + r^2 \dd \Omega^2.
\label{metric-u-v-prime}
\end{align}
We can now introduce the variables $\mathcal{T}$ and $\mathcal{R}$ via the relations
\begin{align}
\mathcal{T} &= \frac{1}{2 \sqrt{\varepsilon}} \left(v^\prime + \varepsilon u^\prime \right), 
\nonumber \\  
\mathcal{R} &= \frac{1}{2 } \left(v^\prime - \varepsilon u^\prime \right), 
\end{align}
and hence the sought-after metric in  Kruskal-Szekeres coordinates $(\mathcal{T},\mathcal{R},\theta,\phi)$ reads as
\begin{align}
\dd s^2 =  \frac{32 M^3}{r} {\rm e}^{-r/(2M)} \left( -\varepsilon   \, \dd \mathcal{T}^2 + \dd \mathcal{R}^2 \right)+ r^2 \dd \Omega^2,
\label{metric-KS-coord}    
\end{align}
where $r$ is defined implicitly by
\begin{align} \label{r-KS-coordinates}
-\varepsilon \mathcal{T}^2 + \mathcal{R}^2 = \left(\frac{r}{2M}-1\right) \ee^{r/(2M)}.      
\end{align}
The  pathological behaviour at $r=2M$ has been removed in Eq. \eqref{metric-KS-coord} (or, equivalently, Eq.  \eqref{metric-u-v-prime}). In fact, apart from the singularity at $r=0$, the only pathology affecting the metric \eqref{metric-KS-coord} is due to the fact that it becomes degenerate on the change surface $\Sigma$, where $\varepsilon=0$.

The calculations  performed in this paper are not straightforward in Kruskal-Szekeres coordinates as the variable $r$ is specified by Eq. \eqref{r-KS-coordinates}. For this reason, in Sec. \ref{Sec:PG-coordinates},  we  used the Gullstrand-Painlev\'e coordinates.

\section{The regularized Riemann tensor in Gullstrand-Painlev\'e coordinates} \label{App:Riemann-tensor}

Starting from the formula \eqref{metric-PG-coordinates} pertaining to the Lorentzian-Euclidean Schwarzschild metric in Gullstrand-Painlev\'e coordinates $(\mathscr{T},r,\theta,\phi)$,  we find that the nonvanishing  components of the Riemann tensor are 
\begin{subequations}
\label{Riemann-tensor-general-expressions}
\begin{align}
R^{r}_{\;r \mathscr{T}r} &= \sqrt{\frac{M}{r}} \frac{ r^2 (2 M-r) \varepsilon^{\prime 2} +2 r \varepsilon  \left[r (r-2 M) \varepsilon^{\prime \prime} +3 M \varepsilon^\prime \right]-8 M \varepsilon^2 }{2 \sqrt{2} r^3 \varepsilon^{3/2}},   
\label{Riemann-1}
\\
R^{r}_{\;\theta \theta r} &= \frac{M}{r},
\\
R^{r}_{\; \phi \phi r} &= \sin^2 \theta\, R^{r}_{\;\theta \theta r},
\\
R^{r}_{\; \mathscr{T} \mathscr{T}r} &= \sqrt{\frac{\varepsilon}{2Mr}}\left(2M-r\right) R^{r}_{\;r \mathscr{T}r},
\label{Riemann-2}
\\
R^{\theta}_{\; r \theta r} &= M \frac{r \varepsilon^\prime - \varepsilon}{r^3 \varepsilon},
\\
R^{\theta}_{\; r \mathscr{T} \theta} &= \sqrt{\frac{M}{r}} \frac{ r (r-2 M) \varepsilon^\prime +2 M \varepsilon }{  \sqrt{2\varepsilon }r^3},
\\
R^{\theta}_{\; \phi \phi \theta} &= -2 \sin^2 \theta \, R^{r}_{\;\theta \theta r} ,
\\
R^{\theta}_{\; \mathscr{T} \mathscr{T} \theta } &= \sqrt{\frac{2 \varepsilon r}{M}}\frac{2M-r}{2r}R^{\theta}_{\; r \mathscr{T} \theta},
\\
R^{\phi}_{\; r \phi r}&= R^{\theta}_{\; r \theta r}, 
\\
R^{\phi}_{\; r \mathscr{T} \phi }&= R^{\theta}_{\; r \mathscr{T} \theta },
\\
R^{\phi}_{\; \theta  \phi \theta }&= 2 R^{r}_{\; \theta \theta r  },
\\
R^{\phi}_{\; \mathscr{T}  \phi r }&= - R^{\theta}_{\; r \mathscr{T} \theta },
\\
R^{\phi}_{\; \mathscr{T} \mathscr{T}  \phi  }&= R^{\theta}_{\; \mathscr{T}\mathscr{T} \theta},
\\
R^{\mathscr{T}}_{\;r  \mathscr{T} r  }&= - \sqrt{\frac{r}{2M\varepsilon}} R^{r}_{\; r \mathscr{T}r},
\label{Riemann-3}
\\
R^{\mathscr{T}}_{\; \theta \theta  r  }&=-\sqrt{\frac{M}{2r}} \frac{r \varepsilon^\prime}{\varepsilon^{3/2}},
\\
R^{\mathscr{T}}_{\; \theta \mathscr{T} \theta    }&= -\frac{M}{r} + (2M-r) \frac{\varepsilon^\prime}{2 \varepsilon},
\\
R^{\mathscr{T}}_{\; \phi \phi r    }&= \sin^2 \theta \;  R^{\mathscr{T}}_{\; \theta \theta   r },
\\
R^{\mathscr{T}}_{\; \phi \mathscr{T} \phi    }&= -\frac{r^2 \sin^2 \theta}{\sqrt{\varepsilon}} \sqrt{\frac{r}{2M}} R^{\theta}_{\; r  \mathscr{T} \theta    },
\\
R^{\mathscr{T}}_{\;  \mathscr{T}  \mathscr{T} r    }&= - R^{r}_{\; r \mathscr{T}  r    },
\label{Riemann-4}
\end{align}    
\end{subequations}
where $\varepsilon(r)$ is given in Eq. \eqref{epsilon-of-r} and its derivatives with respect to $r$ variable are indicated with a prime; the remaining components of the curvature tensor are derived from the  above ones by symmetry.

The factors involving $\varepsilon^{\prime }$ generally give ill-defined expressions which depend linearly on the Dirac delta function $\delta\left(r-2M\right)$. Additionally, in Eqs. \eqref{Riemann-1}, \eqref{Riemann-2}, \eqref{Riemann-3}, and \eqref{Riemann-4},  $\varepsilon^{\prime 2}$  yields indefinite quadratic-in-delta quantities. Ill-defined  $\delta$ and $\delta^2$ terms stem also from $\varepsilon^{\prime \prime}$ pieces in all the aforementioned components except for Eq. \eqref{Riemann-2}, where the $\varepsilon^{\prime \prime}$ contribution results in the integral $\int  \dd r  \left(r-2M\right)^2 \delta^{\prime} \left(r-2M\right)$, which vanishes  by applying conventional distribution-theory techniques. 

The ill-defined entities occurring in Eq. \eqref{Riemann-tensor-general-expressions}  can be regularized by resorting to the scheme introduced in Sec. \ref{Sec:surface-layer}, which permits to obtain the following expression for $R^\alpha_{\; \beta \mu \nu}$ having no distributional behaviour: 
\begin{subequations}
\label{Riemann-regularized}
\begin{align}
R^{r}_{\;r \mathscr{T}r} &= -2 \sqrt{2} \left(\frac{M}{r}\right)^{3/2} \frac{\sqrt{\varepsilon}}{r^2},   
\\
R^{r}_{\;\theta \theta r} &= \frac{M}{r},
\\
R^{r}_{\; \phi \phi r} &= \sin^2 \theta \, R^{r}_{\;\theta \theta r},
\\
R^{r}_{\; \mathscr{T} \mathscr{T}r} &= \frac{2M \varepsilon (r-2M)}{r^4},
\\
R^{\theta}_{\; r \theta r} &= -\frac{1}{r^2} R^{r}_{\;\theta \theta r},
\\
R^{\theta}_{\; r \mathscr{T} \theta} &= -\frac{1}{2}R^{r}_{\;r \mathscr{T}r},
\\
R^{\theta}_{\; \phi \phi \theta} &= -2 \sin^2 \theta \, R^{r}_{\;\theta \theta r},
\\
R^{\theta}_{\; \mathscr{T}  \theta r} &=\frac{1}{2}R^{r}_{\;r \mathscr{T}r},
\\
R^{\theta}_{\; \mathscr{T} \mathscr{T} \theta } &= -\frac{1}{2}R^{r}_{\;  \mathscr{T}  \mathscr{T} r},
\\
R^{\phi}_{\; r \phi r}&= -\frac{1}{r^2} R^{r}_{\;  \theta \theta r}, 
\\
R^{\phi}_{\; r \mathscr{T} \phi }&= -\frac{1}{2} R^{r}_{\; r \mathscr{T} r},
\\
R^{\phi}_{\; \theta  \phi \theta }&= 2 R^{r}_{\; \theta \theta r  },
\\
R^{\phi}_{\; \mathscr{T}  \phi r }&= \frac{1}{2} R^{r}_{\; r \mathscr{T} r },
\\
R^{\phi}_{\; \mathscr{T} \mathscr{T}  \phi  }&= -\frac{1}{2} R^{r}_{\; \mathscr{T}\mathscr{T} r},
\\
R^{\mathscr{T}}_{\;r  \mathscr{T} r  }&= \frac{2}{r^2} R^{r}_{\; \theta \theta r},
\\
R^{\mathscr{T}}_{\; \theta \mathscr{T} \theta    }&= -R^{r}_{\; \theta \theta r},
\\
R^{\mathscr{T}}_{\; \phi \mathscr{T} \phi    }&= -\sin^2 \theta R^{r}_{\;\theta \theta r },
\\
R^{\mathscr{T}}_{\;  \mathscr{T}  \mathscr{T} r    }&= - R^{r}_{\; r \mathscr{T}  r    }.
\end{align}    
\end{subequations}

Although the regularized Riemann tensor \eqref{Riemann-regularized} is discontinuous  across $\Sigma$, it is easy to see that it yields a perfectly well-behaved expression for both the Ricci tensor and  Ricci scalar, which vanish in the whole spacetime manifold as in the standard Lorentzian-signature Schwarzschild solution.

\bibliography{references}

\end{document}